\documentclass[longauth]{aa}
\usepackage[varg]{txfonts}

\usepackage[T1]{fontenc}
\usepackage{ae,aecompl}
\usepackage{multirow}

\usepackage{graphicx}	
\usepackage{amsmath}	
\usepackage{amssymb}	
\usepackage{color}
\usepackage{hyperref}

\begin{document}


\title{The LOFAR Two-metre Sky Survey\thanks{LOTSS}} 

\subtitle{IV. First Data Release: Photometric redshifts and rest-frame magnitudes} 
\authorrunning{Duncan et~al.}
\titlerunning{LoTSS DR1-IV}

\author{Kenneth J Duncan\thanks{E-mail: duncan@strw.leidenuniv.nl}\inst{1}\and
J. Sabater\inst{2}\and  
H. J. A. R\"{o}ttgering\inst{1}\and
M. J. Jarvis\inst{3}\and
D. J. B. Smith\inst{4}\and
P. N. Best\inst{2}\and
J. R. Callingham\inst{5}\and
R. Cochrane\inst{2}\and
J. H. Croston\inst{6}\and
M. J. Hardcastle\inst{4}\and 
B. Mingo\inst{6}\and
L. Morabito\inst{3}\and
D. Nisbet\inst{2}\and
I. Prandoni\inst{7}\and
T. W. Shimwell\inst{5,1}\and
C. Tasse\inst{8}\and
G. J. White\inst{6,9}\and
W. L. Williams\inst{4}\and
L. Alegre\inst{2}\and
K. T. Chy\.zy\inst{10}\and
G. G\"{u}rkan\inst{11}\and
M. Hoeft\inst{12}\and
R. Kondapally\inst{2}\and
A. P. Mechev\inst{1}\and
G. K. Miley\inst{1}\and
D. J. Schwarz\inst{13}\and
R. J. van Weeren\inst{1}\\}
 
\institute{Leiden Observatory, Leiden University, PO Box 9513, NL-2300 RA Leiden, The Netherlands \and
SUPA, Institute for Astronomy, Royal Observatory, Blackford Hill, Edinburgh, EH9 3HJ, UK \and
Astrophysics, University of Oxford, Denys Wilkinson Building, Keble Road, Oxford, OX1 3RH, England \and
Centre for Astrophysics Research, School of Physics, Astronomy and Mathematics, University of Hertfordshire, College Lane, Hatfield AL10 9AB, UK \and
ASTRON, the Netherlands Institute for Radio Astronomy, Postbus 2, 7990 AA, Dwingeloo, The Netherlands \and
School of Physical Sciences, The Open University, Walton Hall, Milton Keynes, MK7 6AA, UK \and
INAF - Istituto di Radioastronomia, via Gobetti 101, 40129 Bologna, Italy \and
GEPI, Observatoire de Paris, CNRS, Universit\'{e} Paris Diderot, 5 place Jules Janssen, 92190 Meudon, France \and
Space Science \& Technology Department, The Rutherford Appleton Laboratory, Chilton, Didcot, Oxfordshire OX11 0NL, UK \and
Astronomical Observatory, Jagiellonian University, ul. Orla 171, 30-244 Krak\'ow, Poland\and
CSIRO Astronomy and Space Science, PO Box 1130, Bentley WA 6102, Australia\and
Thüringer Landessternwarte (TLS), Sternwarte 5, D-07778 Tautenburg, Germany\and
Fakult\"{a}t f\"{u}r Physik, Universit\"{a}t Bielefeld, Postfach 100131, 33501 Bielefeld, Germany
}

\date{Received 4 June 2018 / Accepted 6 August 2018}

\abstract{The LOFAR Two-metre Sky Survey (LoTSS) is a sensitive, high-resolution 120-168\,MHz survey of the Northern sky. 
The LoTSS First Data Release (DR1) presents 424 square degrees of radio continuum observations over the HETDEX Spring Field (10h45m00s $<$ right ascension $< $ 15h30m00s and 45$^\circ$00$\arcmin$00$\arcsec$ $<$ declination $<$ 57$^\circ$00$\arcmin$00$\arcsec$) with a median sensitivity of 71$\mu$Jy/beam and a resolution of 6$\arcsec$.
In this paper we present photometric redshifts (photo-$z$) for 94.4\% of optical sources over this region that are detected in the Panoramic Survey Telescope and Rapid Response System (Pan-STARRS) 3$\pi$ steradian survey.
Combining the Pan-STARRS optical data with mid-infrared photometry from the Wide-field Infrared Survey Explorer, we estimate photo-$z$s using a novel hybrid photometric redshift methodology optimised to produce the best possible performance for the diverse sample of radio continuum selected sources. For the radio-continuum detected population, we find an overall scatter in the photo-$z$ of 3.9\% and an outlier fraction ($\left |  z_{\textup{phot}} - z_{\textup{spec}} \right | / (1+z_{\text{spec}}) > 0.15$) of $7.9\%$.
We also find that, at a given redshift, there is no strong trend in photo-$z$ quality as a function of radio luminosity.
However there are strong trends as a function of redshift for a given radio luminosity, a result of selection effects in the spectroscopic sample and/or intrinsic evolution within the radio source population.
Additionally, for the sample of sources in the LoTSS First Data Release with optical counterparts, we present rest-frame optical and mid-infrared magnitudes based on template fits to the consensus photometric (or spectroscopic when available) redshift.} 

\maketitle 

\defcitealias{Duncan:2017wu}{D18a}
\defcitealias{Duncan:2017ul}{D18b}
\defcitealias{Shimwell:2018to}{DR1-I}
\defcitealias{Williams:2018us}{DR1-II}


\section{Introduction}
With its exquisite sensitivity and excellent field-of-view, the Low Frequency Array \citep[LOFAR;][]{vanHaarlem:2013gi} is a powerful new tool for deep radio continuum surveys.
The LOFAR Two-metre Sky Survey (LoTSS) is currently undertaking a survey of the northern sky at 120-168MHz\footnote{Formally, the central frequency of the LoTSS first data release is 144MHz. However, throughout this paper we will refer to the LoTSS frequency colloquially as 150 MHz}.
In the first release of data to the full intended depth and angular resolution, the first paper in this series \citet[][DR1-I hereafter]{Shimwell:2018to} present observations of over 400 square degrees of the Hobby-Eberly Telescope Dark Energy Experiment (HETDEX) Spring Field (over the region 10h45m00s $<$ right ascension $< $ 15h30m00s and 45$^\circ$00$\arcmin$00$\arcsec$ $<$ declination $<$ 57$^\circ$00$\arcmin$00$\arcsec$).
Reaching a median sensitivity of 71$\mu$Jy/beam with a resolution of $\sim6\arcsec$, the resulting radio continuum catalog consists of over 318,000 sources.

Extracting the maximum scientific data from the LoTSS data firstly requires robust identification of the host-galaxies of radio sources.
Secondly, we require knowledge of the source redshifts to extract intrinsic physical properties for both the radio sources (e.g. physical size, luminosity) and their host galaxies.
In the second paper in this series, \citet[][DR1-II hereafter]{Williams:2018us} present details of the extensive optical cross-matching procedure used to identify counterparts within the available all-sky optical (and mid-infrared) photometric surveys. 
In this paper, we present redshift estimates for both the corresponding optical and radio sources as well as estimates of the radio host-galaxy rest-frame optical properties - providing the community with a value-added catalog that can enable a wide variety of radio continuum science.
 
Future spectroscopic surveys such as WEAVE-LOFAR \citep{Smith:2016vw} will provide precise redshift estimates and robust source classification for large numbers of the LoTSS source population.
Using the WHT Enhanced Area Velocity Explorer \citep[WEAVE; ][]{Dalton:2012hz} $\sim1000$ fibre multi-object spectrograph, WEAVE-LOFAR will obtain $>10^{6}$ spectra for radio sources from the LOFAR 150 MHz survey.
However, WEAVE-LOFAR will only target a small fraction ($\lesssim 5\%$) of the $\gtrsim 15$ million radio sources LoTSS is expected to detect.
Accurate and unbiased photometric redshift estimates for the remaining radio sources will therefore be essential for maximising the scientific potential of LoTSS.

A potential difficulty in estimating photo-$z$s for the radio continuum population is that it is extremely diverse - with synchrotron radio emission tracing both a range of phases of black hole accretion in AGN and star formation activity.
Photo-$z$ techniques optimised for one subset of the radio population (e.g. for star-forming galaxies or for luminous quasars with problematic high equivalent width (EW) emission lines) will produce poor results for the other populations.
Furthermore, in many cases we do not necessarily know \emph{a priori} the nature of a given radio source and therefore the optimum method to apply.
In two recent works, \citet[][hereafter D18a]{Duncan:2017wu} and \citet[][hereafter D18b]{Duncan:2017ul} we have developed and tested a novel photo-$z$ method designed to produce the best possible photo-$z$ estimates for all galaxy types.
By combining multiple estimates, including both traditional template fitting and empirical training based (or `machine learning') methods, it is possible to produce a consensus redshift estimate that combines the strengths of the different techniques.
In this paper we detail how this method was applied to the optical data in the LoTSS Data Release region, explore the accuracy of the resulting estimates with respect to key optical and radio properties and present rest-frame optical properties derived from these redshifts.

This paper is organized as follows: In Section~\ref{sec:data} we summarise the data used for estimating photo-$z$s, including the input photometry, multi-wavelength classifications using external optical and X-ray information and details of the spectroscopic training and test sample.
In Section~\ref{sec:photozmethod} we outline the photo-$z$ methodology as implemented for this specific work, detailing key differences from the deep field analysis presented in \citetalias{Duncan:2017wu} and \citetalias{Duncan:2017ul}.
In Section~\ref{sec:results} we analyse the precision and accuracy of the resulting photo-$z$ as a function of key properties - particularly for the LOFAR radio continuum selected population.
In Section~\ref{sec:derivedquantities} we present details of additional rest-frame properties calculated using the derived photo-$z$s.
In Section~\ref{sec:catalog} we provide a description of the final photo-$z$ catalog and the columns it includes.
Finally in Section~\ref{sec:summary} we present a summary of our work.

Throughout this paper, all magnitudes are quoted in the AB system \citep{1983ApJ...266..713O} unless otherwise stated. We also assume a $\Lambda$-CDM cosmology with $H_{0} = 70$ kms$^{-1}$Mpc$^{-1}$, $\Omega_{m}=0.3$ and $\Omega_{\Lambda}=0.7$.

\section{Data}\label{sec:data}
\subsection{Photometry}
In this work we estimate photo-$z$s using the catalogs presented in \citetalias{Williams:2018us} for the optical cross-identification.
Here we outline specific reasons for the choice of photometry used and details of additional processing that was done before photo-$z$ analysis.
To maximise the available information for faint sources, we make use of the forced photometry columns in the PanSTARRs database - specifically the forced aperture photometry columns (FApFlux and FApFluxErr).
The key benefit of the forced photometry values over the default PS1 photometry is that flux information is available in the case of non-detections.
When estimating photo-$z$s for high redshift sources, flux measurements of such non-detections are crucial in accurately constraining, for example, the Lyman break feature.
		
Mid-infrared photometry is taken from the Wide-field Infrared Survey Explorer mission \citep[WISE;][]{Wright:2010in}. 
Specifically we use the AllWISE photometry that combines data from the cryogenic and post-cryogenic \citep[NEOWISE;][]{2011ApJ...731...53M} missions.
The WISE profile-fitting magnitudes and corresponding uncertainties are converted to AB magnitudes and $\mu$Jy flux units consistent with the optical photometry following the prescription outlined in the All-Sky Data Release Explanatory Supplement \footnote{\href{http://wise2.ipac.caltech.edu/docs/release/allsky}{All-Sky Data Release Explanatory Supplement: http://wise2.ipac.caltech.edu/docs/release/allsky}}.
An additional flux uncertainty of 10\% is also added in quadrature to the W3 and W4 flux uncertainties following the recommendations in the Explanatory Supplement (this uncertainty is mainly due to discrepancies in calibrators used for WISE).

Finally, before any training or template-fitting, we correct all optical/mid-infrared photometry values for galactic extinction.
Estimates of $E(B-V)$ for each source position are from \citet{1998ApJ...500..525S}, queried through the Argonaut API \citep{2015ApJ...810...25G}\footnote{\href{http://argonaut.skymaps.info}{Argonaut API:http://argonaut.skymaps.info}}.
Filter-dependent extinction factors are then calculated by convolving the respective filter response curves with the Milky Way dust extinction law of \citet{1999PASP..111...63F}.

Additional details on optical catalogs, including the cross-matching procedure used to join the PanSTARRs and WISE catalogs, are outlined in the companion paper, \citetalias{Williams:2018us}.

\subsection{Multi-wavelength Classifications}\label{sec:mw_class}

As in \citetalias{Duncan:2017wu} and \citetalias{Duncan:2017ul}, for the purposes of optimising photo-$z$ estimates for different subsets of the optical (and radio) population we identify known optical, X-ray and infrared AGN candidates within the optical source catalog.

\begin{itemize}
\item \emph{Optical AGN} are identified primarily through cross-matching of the optical catalogs with the Million Quasar Catalog compilation of optical AGN, primarily based on SDSS  \citep{2015ApJS..219...12A} and other literature catalogs \citep{2015PASA...32...10F}. 
Sources which have been spectroscopically classified as AGN are also flagged.
Objects in the million quasar catalog were cross-matched to the photometric catalogs using a simple nearest neighbour match in RA and declination and allowing a maximum separation of 1\arcsec.

\item Bright \emph{X-ray} sources in the HETDEX field were identified based on the Second Rosat all-sky survey \citep[2RXS;][]{2016A&A...588A.103B} and the XMM-Newton slew survey (XMMSL2)\footnote{\hyperlink{https://www.cosmos.esa.int/web/xmm-newton/xmmsl2-ug}{https://www.cosmos.esa.int/web/xmm-newton/xmmsl2-ug}}.
X-ray sources were matched to their optical counterparts using the published AllWISE cross-matches of \citet{Salvato:2017gj}.
Details of the novel statistical cross-matching code, \textsc{NWay}, used to identify counterparts for the imprecise X-ray source positions are presented in \citet{Salvato:2017gj}.
Matching from the published AllWISE counterparts to the combined HETDEX photometric dataset was done using the corresponding AllWISE source positions. 
 
Following the additional X-ray to WISE colour criteria presented in \citet{Salvato:2017gj}, we additionally separate the AGN and star-forming (or stellar) X-ray source populations such that for AGN:
\begin{equation}
 [W1] > -1.625 \times \log_{10}(F_{0.5-2\textup{keV}}) -8.8	,
\end{equation}
where $[W1]$ is the AllWISE W1 magnitude \emph{in Vega} magnitudes and $F_{0.5-2\textup{keV}}$ the 2RXS or XMMSL2 flux in units of erg$^{-1}$~s$^{-1}$~cm$^{-2}$.
Based on this classification, we define the `XrayClass' as 0 for sources with no X-ray detection, 1 for X-ray sources classified as AGN and 2 for X-ray sources classified as galaxies or stars.

\item \emph{Infrared AGN} are identified using the WISE mid-infrared photometry. \citet{2013ApJ...772...26A} present a range of colour (and magnitude) based selection criteria using the W1 and W2 bands that are designed to select mid-infrared AGN at 75 and 90\% completeness and `reliability', labelled $C_{75}$/ $C_{90}$ and $R_{75}$/$R_{90}$ respectively.
	For every WISE-detected source in the full photometric catalog, we apply all four selections in order of increasing strictness to produce a binary flag, `IRClass', that enables easy selection of the desired criteria.
	The order and corresponding flag values are: $C_{90}$ (1) > $C_{75}$ (2) >  $R_{75}$ (4) > $R_{90}$ (8).
	For example, a source which satisfies both completeness criteria and the `75\% reliability' criteria, $R_{75}$, would have an IRClass $= 7$. 
	For the purposes of the photo-$z$ training and estimation, we use the $R_{75}$ criteria (IRClass $> 4$). This selection yields a total of $\sim10^{5}$ sources in the full photometric sample classified as IR AGN.
\end{itemize}

\begin{figure}
  \centering
  \includegraphics[width=0.9\columnwidth]{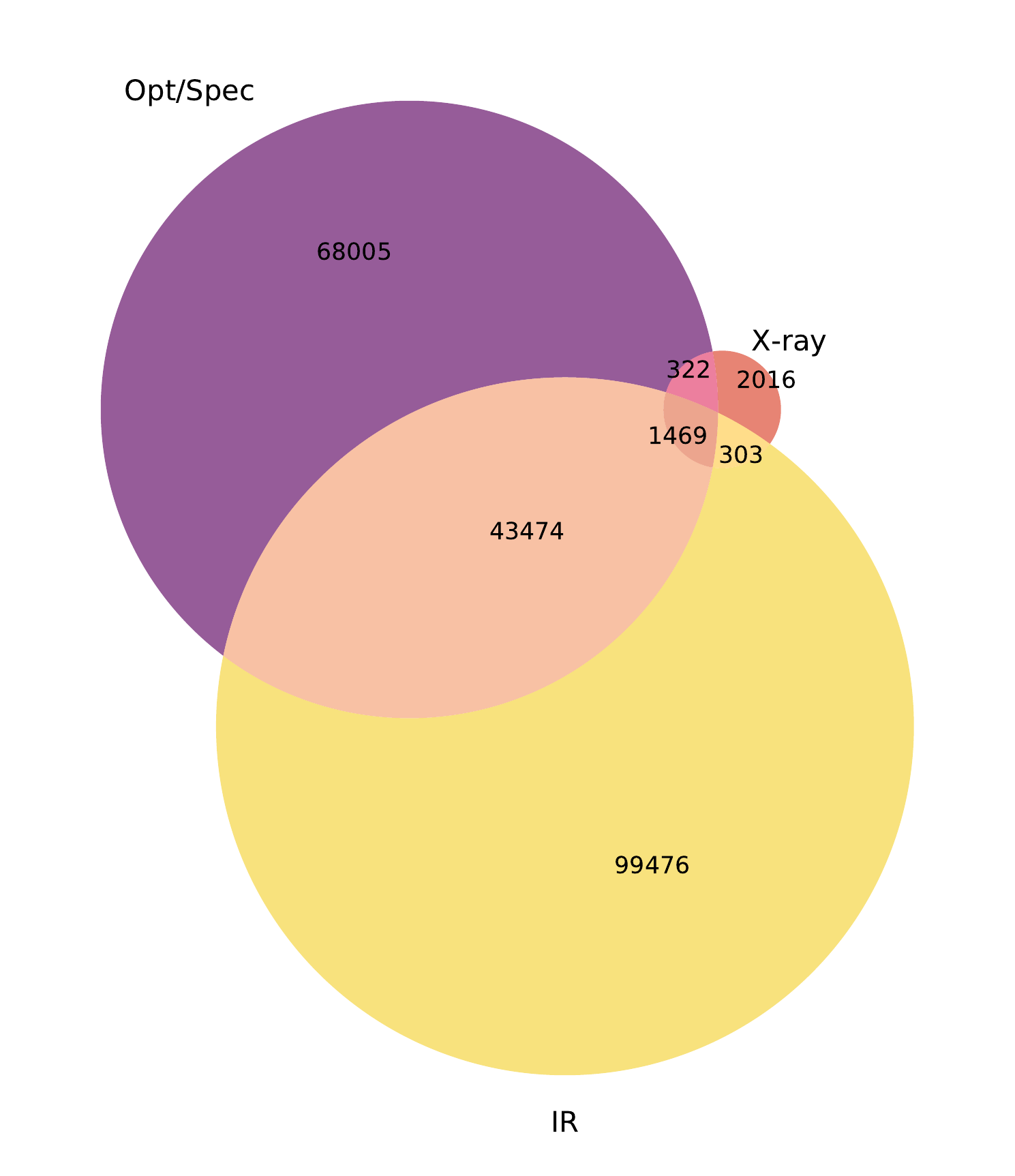}
  \caption{Subsets of the full HETDEX PS1+WISE photometric sample identified as optically, X-ray selected, or infrared selected AGN. Details of the different selection criteria are outlined in the main text. The labelled sample sizes corresponding to a given region do not include subsets of that class, for example 99746 corresponds to the sources are selected as IR AGN but do no pass any other criteria (rather than all IR selected sources).}
  \label{fig:agn_venn_all}
\end{figure}

\begin{figure}
  \centering
  \includegraphics[width=0.9\columnwidth]{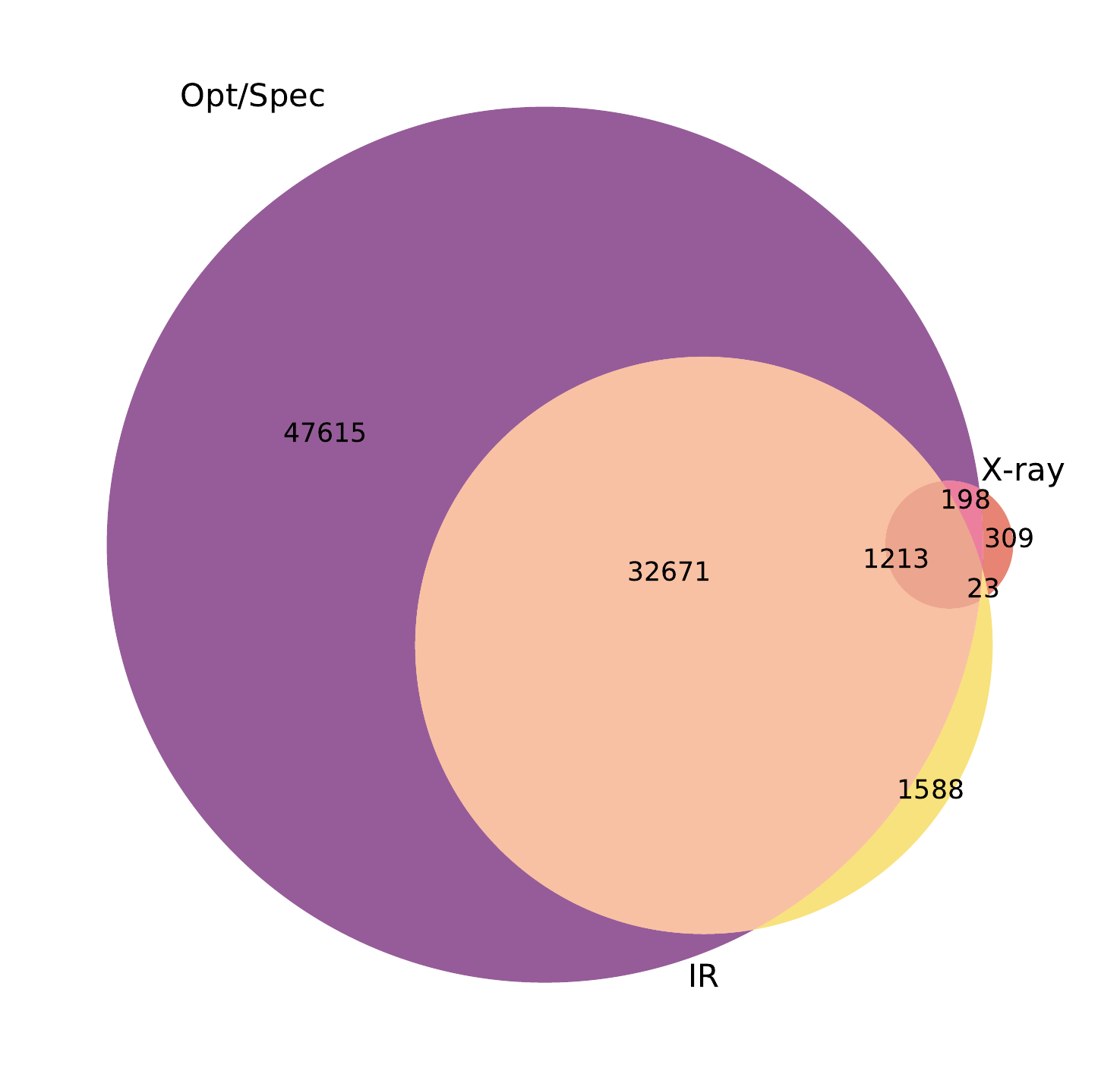}
  \caption{Subsets of the HETDEX, EGS and Bo\"{o}tes PS1+WISE spectroscopic training sample identified as optically, X-ray selected, or infrared selected AGN. Details of the different selection criteria are outlined in the main text. The labelled sample sizes corresponding to a given region do not include subsets of that class, for example 309 corresponds to the sources are selected as X-ray AGN but do no pass any other criteria (rather than all X-ray selected sources).}
  \label{fig:agn_venn_spec}
\end{figure}

In Fig.~\ref{fig:agn_venn_all} and \ref{fig:agn_venn_spec} we illustrate the relative size of each of these subsets within the full photometric and spectroscopic training samples respectively.
As seen in previous work, there is significant overlap between sources selected by each of these criteria.
In comparison with similar criteria applied in fields with significantly deeper targeted X-ray surveys, the relative number of X-ray selected AGN in our sample is very small.
Nevertheless, the scientific potential offered by the deep LOFAR observations of these bright X-ray sources merits their continued inclusion and separate treatment in the rest of our analysis.
All of the AGN selection classes are included within the value-added catalog for convenience.
We note however, that these classifications are not intended to be complete or exhaustive  classifications of AGN source types.

\subsection{Spectroscopic Training and Test Sample}
The majority of spectroscopic redshifts used for training and testing the photo-$z$ in this work are taken from the Sloan Digital Sky Survey Data SDSS Release 14.
For any source in SDSS DR14 classified as a QSO, we use the separate QSO redshift catalog published in \citet{2017arXiv171205029P}.

In addition to the SDSS sources across the full HETDEX field, we include two additional deep spectroscopic training samples
Firstly, we include additional spectroscopic data from the Extended Groth Strip deep field within the wider HETDEX field.
Redshifts in this field are compiled for a range of deep optical surveys within the literature.


Secondly, we include an additional training sample from the $\sim9$ deg$^{2}$ NOAO Deep Wide Field Survey in Bo\"{o}tes \citep[NDWFS:][]{Jannuzi:1999wu} outside the HETDEX footprint.
PanSTARRS and WISE optical catalogs were produced for the field following the same matching procedure as used for the main data sample.
The additional Bo\"{o}tes optical sources were then matched to the available literature spectroscopic redshifts in the field.
In Bo\"{o}tes the bulk of the spectroscopic redshifts are taken from the AGN and Galaxy Evolution Survey \citep[AGES;][]{Kochanek:jy} spectroscopy campaign, with additional samples provided by numerous follow-up surveys in the field including \citet{2012ApJ...758L..31L,2013ApJ...771...25L,2014ApJ...796..126L}, \citet{2012ApJ...753..164S}, \citet{2012ApJ...756..115Z,2013ApJ...779..137Z} and \citet{2016ApJ...823...11D}.

\begin{figure}
  \includegraphics[width=\columnwidth]{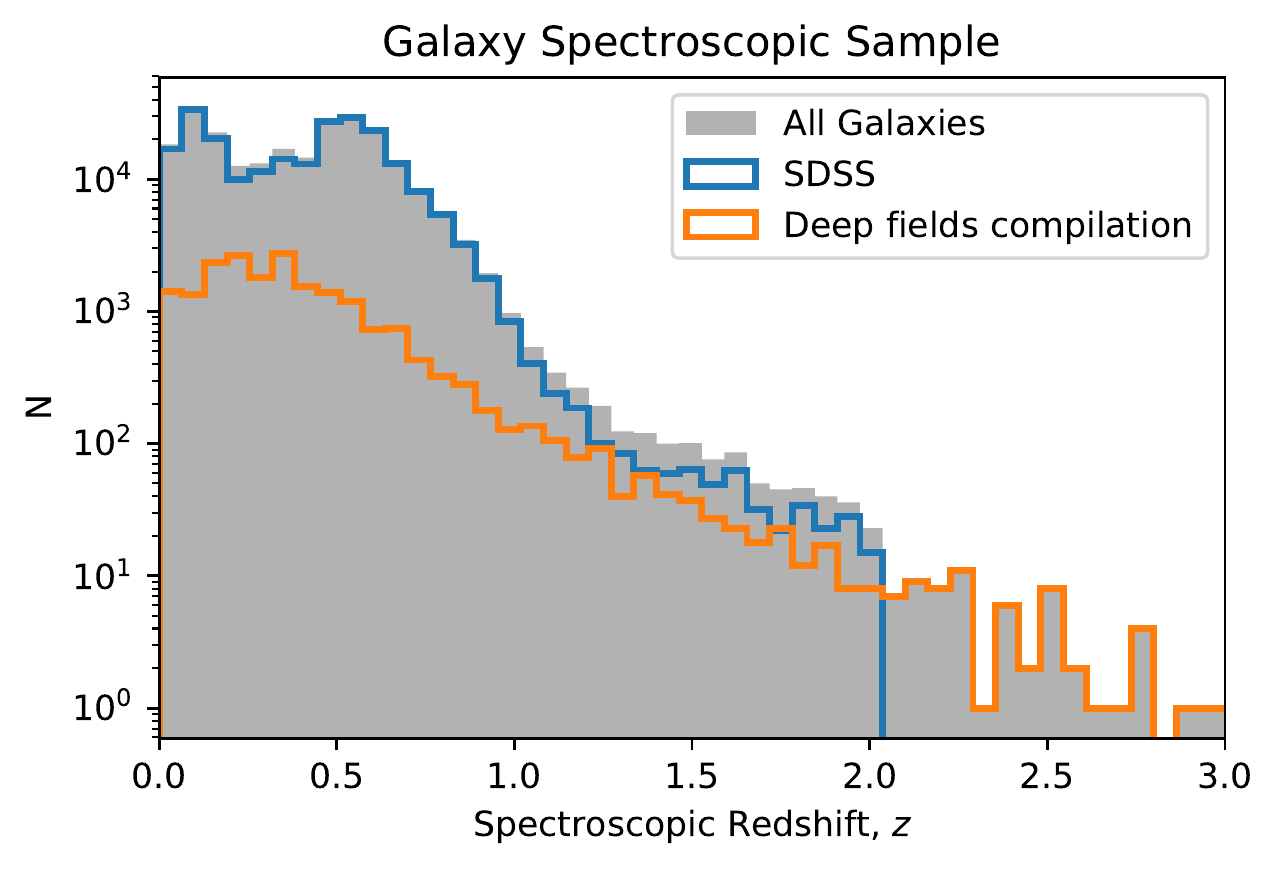}
  \caption{Spectroscopic redshift distribution for the training sample sources that do not satisfy any of the  multi-wavelength AGN selection criterion.}
  \label{fig:specz_hist_gal}
\end{figure}

\begin{figure}
  \includegraphics[width=\columnwidth]{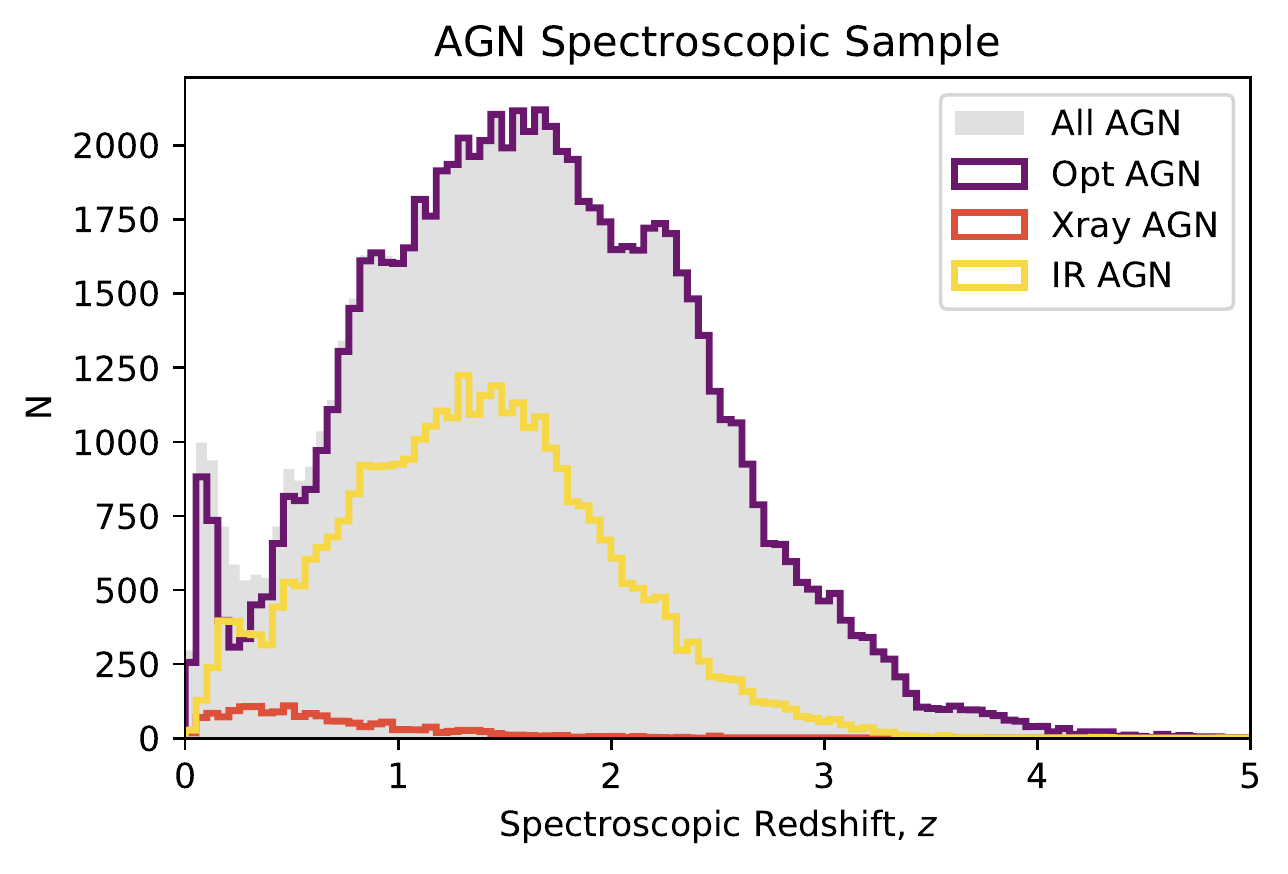}
  \caption{Spectroscopic redshift distribution for the training sample sources identified as optically, X-ray selected, or infrared selected AGN. Note that as illustrated in Fig.~\ref{fig:agn_venn_all}, many sources satisfy more than one multi-wavelength AGN selection criteria.}
    \label{fig:specz_hist_agn}

\end{figure}

In total the combined spectroscopic training sample consists of 336,499 sources, of which 83,617 satisfy one or more of the AGN selection criteria.
For the LOFAR detected sources in this field presented in LoTSS DR1, 29535 sources have spectroscopic redshifts in our current compilation.
Fig.~\ref{fig:specz_hist_gal} and \ref{fig:specz_hist_agn} show histograms of the spectroscopic redshift distribution for the various subsets of the spectroscopic sample.
We note that while the `Deep' sample does increase the available range of redshifts, between $0.5 \lesssim z \lesssim 1$ the training sample falls away very rapidly.
Similarly for the AGN sample, the number of $z_{\textup{spec}}$ available above $z > 3$ is proportionally very small.


\section{Photometric Redshift Methodology}\label{sec:photozmethod}
To estimate photometric redshifts for the complete HETDEX region whilst optimising the performance for the LOFAR detected population, we make use of the hybrid photo-$z$ method that is presented in \citetalias{Duncan:2017wu} and \citetalias{Duncan:2017ul}.
The method is `hybrid' in the sense that it combines both machine learning and template-fitting based photo-$z$ estimates together to produce a combined consensus estimate designed to combine the strengths of each method.
In this section we present a summary of the method and how it was applied to the combined PS1 + WISE photometric dataset within HETDEX.
In Section~\ref{sec:method-gpz} we describe the derivation of our machine learning based redshift estimates, while Section~\ref{sec:method-temp} describes our template photo-$z$ methodology. 
Section 3.3 describes our method of combining these methods to produce an optimised consensus redshift estimate.

\subsection{Gaussian Process Estimates}\label{sec:method-gpz}
The `machine learning' aspect of the hybrid photo-$z$s are produced using the Gaussian process redshift code, \textsc{GPz} \citep{2016MNRAS.455.2387A,2016MNRAS.462..726A}.
\textsc{GPz} models the distribution of functions that map a given set of input vectors, in this case a training set of magnitudes and corresponding uncertainties, onto the desired output, i.e. the spectroscopic redshift.
The trained model can then be used to predict the redshift for a new set of input magnitudes and uncertainties.

Three advantages of \textsc{GPz} over other implementations of Gaussian processes or alternative empirical methods in the literature are: firstly, lower computational requirements without significantly affecting accuracy by introducing a sparse GP framework \citep{2016MNRAS.455.2387A} .
Secondly, by modelling both the intrinsic noise within the photometric data and model uncertainties due to limited training data, \textsc{GPz} is able to account for non-uniform and variable noise (heteroscedastic) within the input data.
Finally, by incorporating so-called `cost-sensitive learning', \textsc{GPz} can optimise the analysis for a specific science goal by giving different weights to different parts of the training sample parameter space.
Further details of the theoretical background and methodology of \textsc{GPz} and their specific implementation can be found in \citet{2016MNRAS.455.2387A} and \citet{2016MNRAS.462..726A}.

\subsubsection{Main galaxy population}
In our current implementation of \textsc{GPz} within the hybrid photo-$z$ framework, we use magnitudes and magnitude errors as the input data.
Sources that have no measurement or negative flux will therefore have no measurements and cannot be included in the training or have redshift predictions.
However, we still wish to maximise the number of sources for which we can produce a \textsc{GPz} estimate whilst also obtaining the best estimate available for a given source.
Therefore, for the sources that do not satisfy any of the multi-wavelength AGN criteria we train three different \textsc{GPz} classifiers on subsets of the data with an increasing number of bands (with a decreasing number of sources then detected in all bands).
We use:
\begin{itemize}
	\item PS1 $r$, $i$ and $z$ - For the PS1 dataset, this combination of bands maximises the number of sources with magnitude measurements in three bands. After restricting the catalogs to sources with magnitudes in these three bands we are left with 98.7\% of the training sample and 74.3\% of the full optical catalog
	\item PS1 $g$, $r$, $i$, $z$ and $y$ - Magnitudes are available for all optical bands in 97\% of the training sample and 66.3\% of the full catalog
	\item PS1 $g$, $r$, $i$, $z$, $y$ and WISE W1 - When including the additional criteria of a WISE detection, there are significantly fewer sources, with 84.2\% of the training sample and just 27.9\% of the full sample. However, as will be shown later the inclusion of WISE in the photo-$z$ estimates yields significant improvement. 
\end{itemize}

\begin{figure}
  \includegraphics[width=\columnwidth]{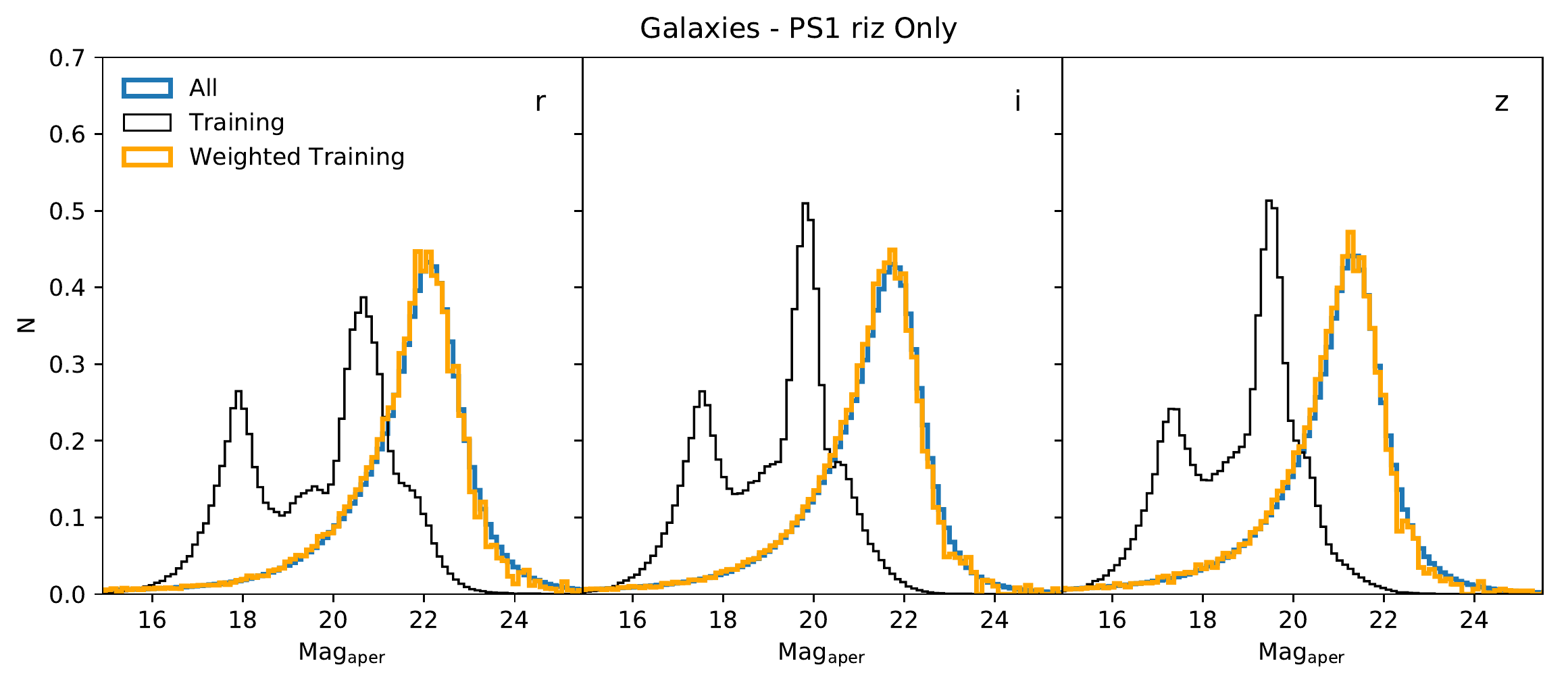}
\includegraphics[width=\columnwidth]{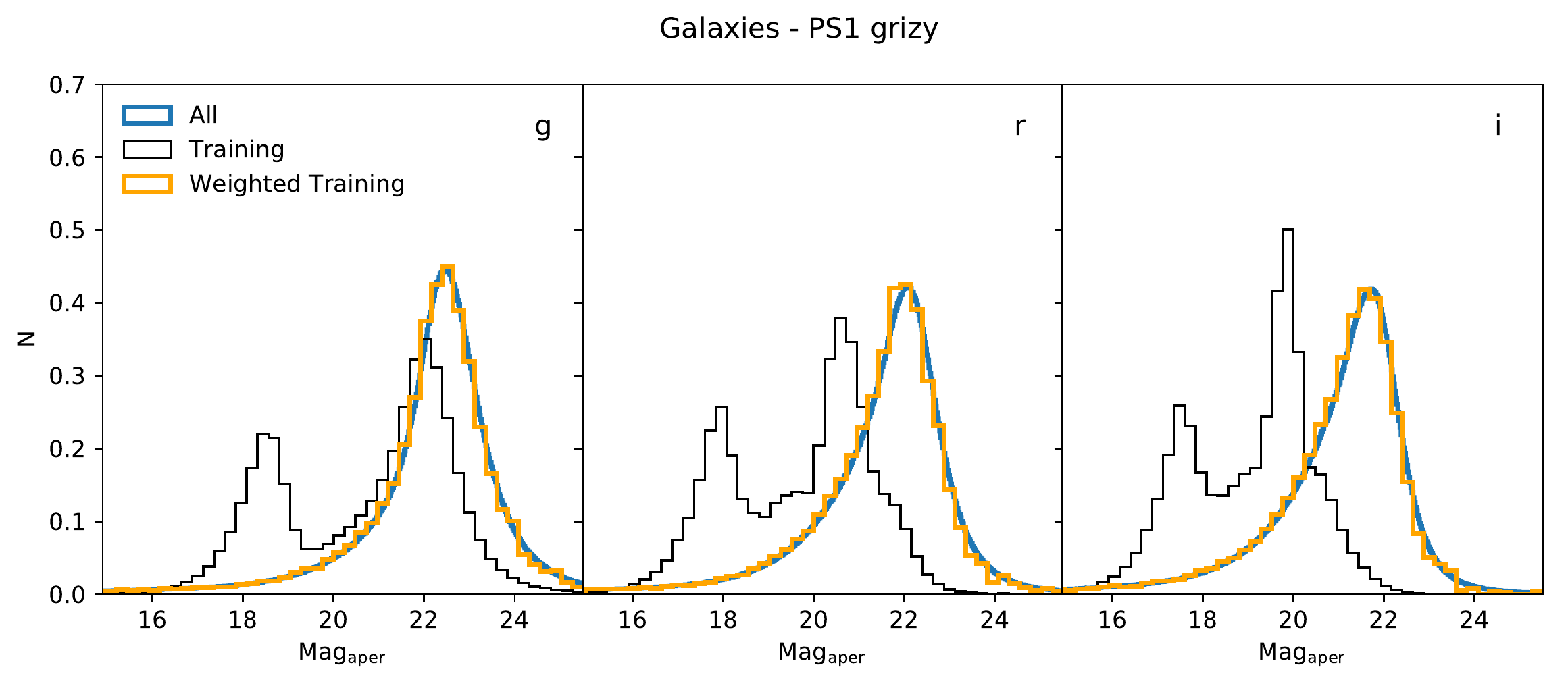}
\includegraphics[width=\columnwidth]{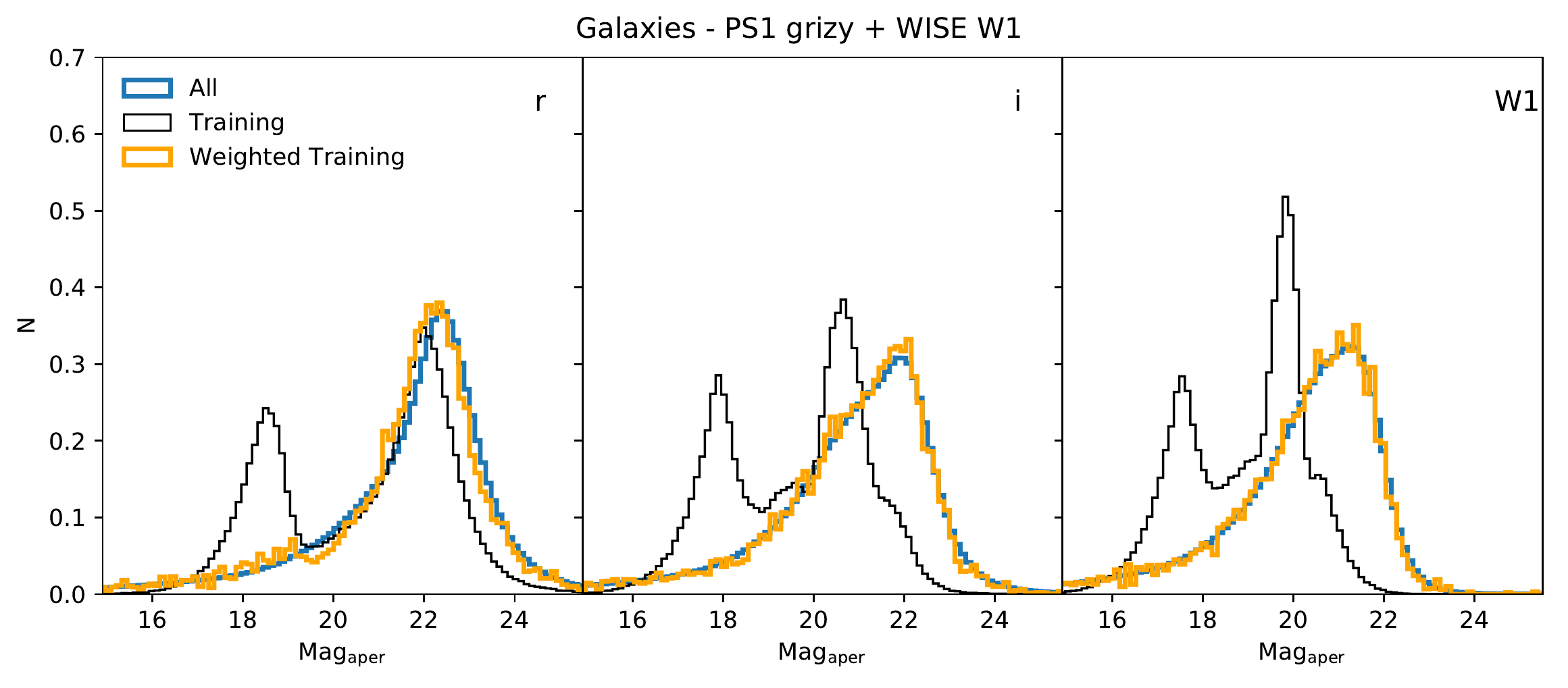}
  \caption{Illustration of the colour-magnitude based weighting scheme applied to the AGN training subsets employed in this work. The thick blue line shows the magnitude distributions for the full photometric sample while the thin black and thick gold lines show the training sample before and after weighting. For each magnitude distribution, the corresponding photometric band is labelled in the upper right corner of the panel.}
  \label{fig:weights_gal}
\end{figure}

When training the \textsc{GPz} classifiers, we employ a weighting scheme based on the method presented in \citet{Lima:2008eu} that takes into account the colour and magnitude distribution of the training sample with respect to the full corresponding photometric sample.
This weighting scheme allows us to account for potential biases in the training sample to produce estimates that are optimised for the bulk of the galaxy population rather than just the bright population with better spectroscopic coverage.

To calculate the weights for each sample we use the $i$-band magnitude plus two additional colours.
For the $riz$-only training sample the additional colours used are $r - i$ and $i-z$, while for the $grizy$ sample we use $g-r$ and $r-i$.
Finally for the $grizy$ + $W1$ sample we use $r-i$ and $i-W1$ colours.
We note however that the specific choice of colours are not critical and the weighting scheme is typically able to closely reproduce the magnitude distribution in all bands regardless of whether they were included in the weight calculations or not.

In Fig.~\ref{fig:weights_gal} we illustrate the results of the weighting scheme for each of the galaxy training samples listed above.
For the three magnitudes used in the weighting scheme, we show the magnitude distribution of the full photometric sample compared to that of the training sample both before and after the weighting scheme has been applied.

As in \citetalias{Duncan:2017ul}, we train \textsc{GPz} using 25 basis functions and allowing variable covariances for each basis function \citep[i.e. the `GPVC' of][]{2016MNRAS.455.2387A}.
Finally, we also follow the practices outlined in Section 6.2 of \citet{2016MNRAS.455.2387A} and allow pre-processing of the input data to normalise or de-correlate the features (also known as `sphering' or `whitening').

\begin{figure*}
\centering
  \includegraphics[width=0.25\paperwidth]{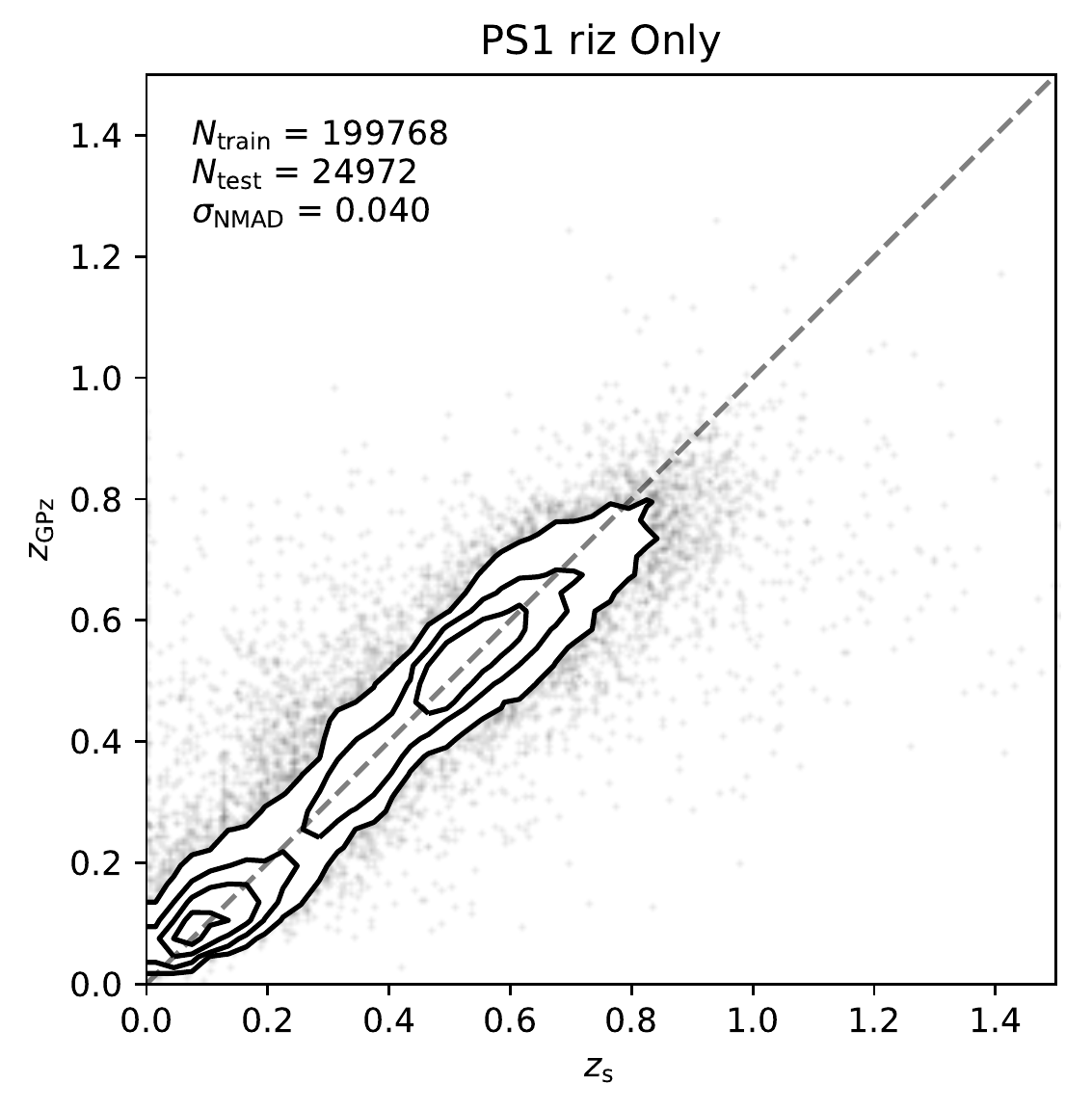}
  \includegraphics[width=0.25\paperwidth]{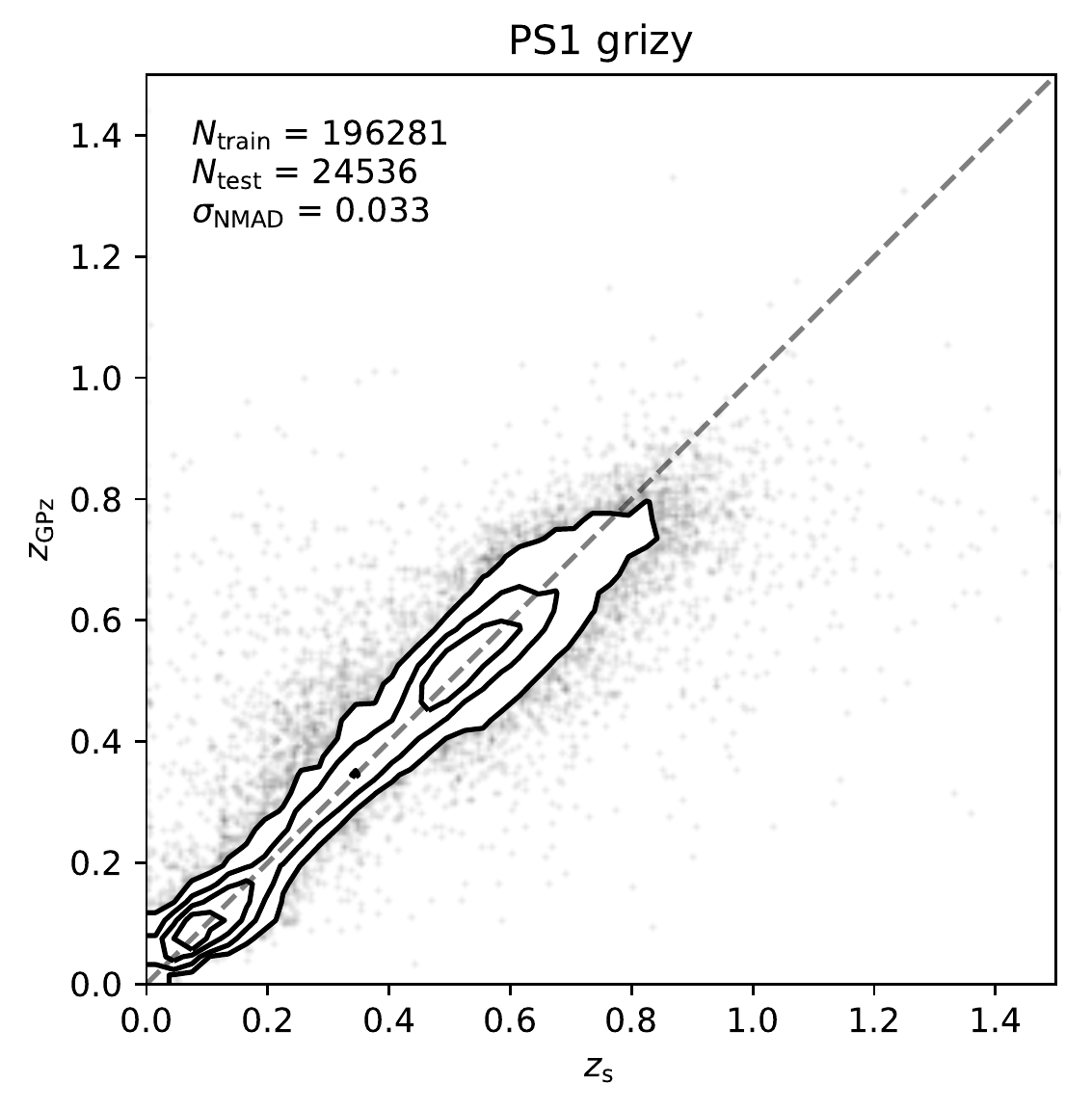}
  \includegraphics[width=0.25\paperwidth]{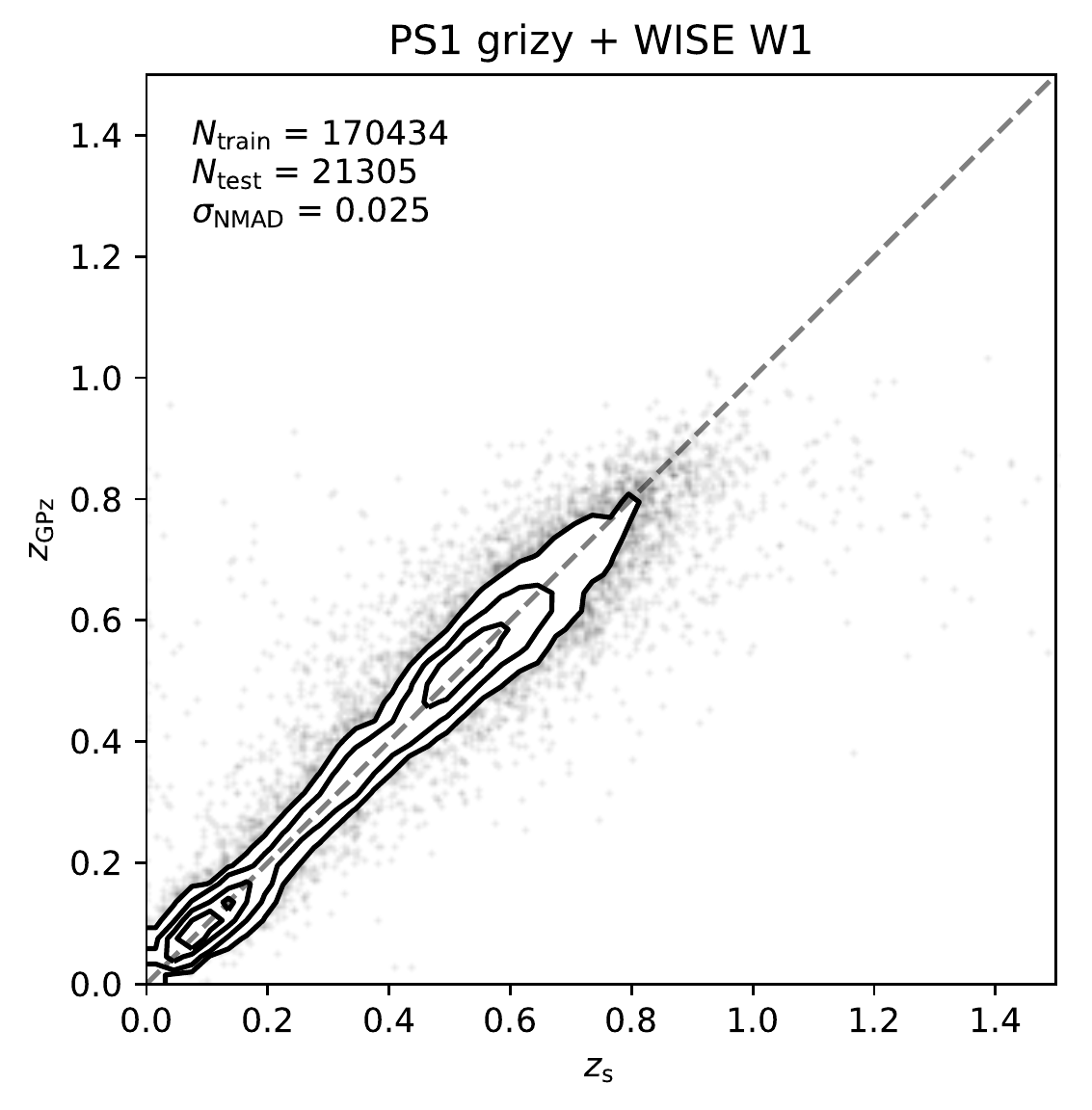}
  \caption{Distribution of \textsc{GPz} photometric redshift estimates vs spectroscopic redshift for the galaxy test sample (not included in training in any way) for the three different detection criteria. The number of training sources used ($N_{\textup{train}}$), the number of test sources plotted ($N_{\textup{test}}$) and the corresponding robust scatter for the test sample ($\sigma_{\textup{NMAD}}$) are shown in the the upper left corner of each panel. The plotted contours are linearly spaced in source density.}
  \label{fig:gpz_gal_speczphotz}
\end{figure*}

Fig.~\ref{fig:gpz_gal_speczphotz} presents the resulting photo-$z$ quality of the \textsc{GPz} spectroscopic test sample (10\% of the training sample subset not included in the \textsc{GPz} training or validation in any way) for each of the training samples.
Based on the density contours it is evident that the \textsc{GPz} photo-$z$ performance for galaxies is excellent in all of the training samples out to redshifts of $z \approx 0.8$. 
Above this redshift, the training sample becomes particularly sparse (see Fig.~\ref{fig:specz_hist_gal}) and the estimates become increasingly biased.
Quantitatively, the overall scatter for the \textsc{GPz} redshifts ranges from 4\% ($riz$) to 2.5\% for sources with WISE W1 detections.
More detailed quantitative analysis of the photo-$z$ quality is reserved for the final hybrid estimates.

\subsubsection{Optical, X-ray and Infrared selected AGN subsets}
\textsc{GPz} photo-$z$ estimates for sources that satisfy any of the additional multi-wavelength AGN criteria are produced for training samples based on the optically (quasar), X-ray and infrared subsets.
For all three subsets, we make use of full set of PS1 optical bands ($g$, $r$, $i$, $z$ and $y$) as well as WISE W1 $3.6\mu$m band.

\begin{figure}
\centering
  \includegraphics[width=1\columnwidth]{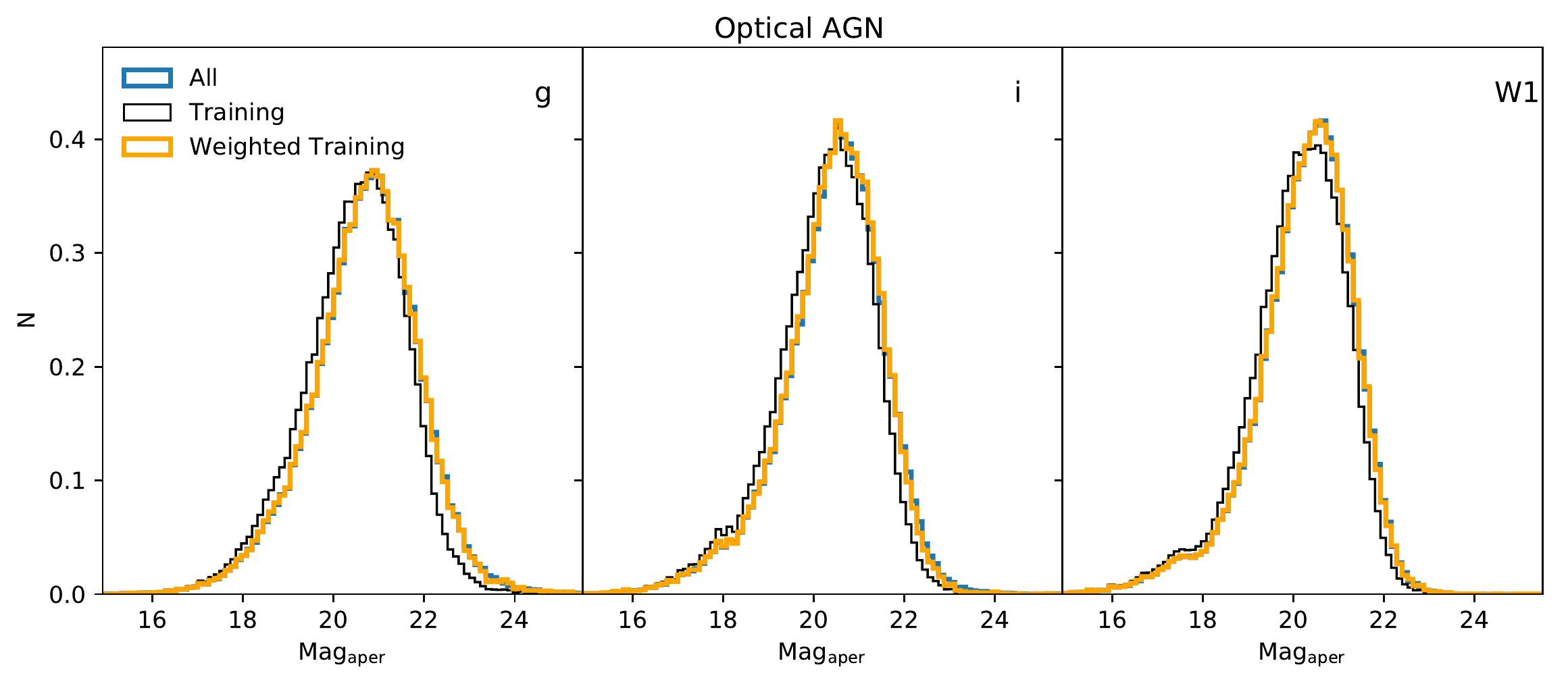}
  \includegraphics[width=1\columnwidth]{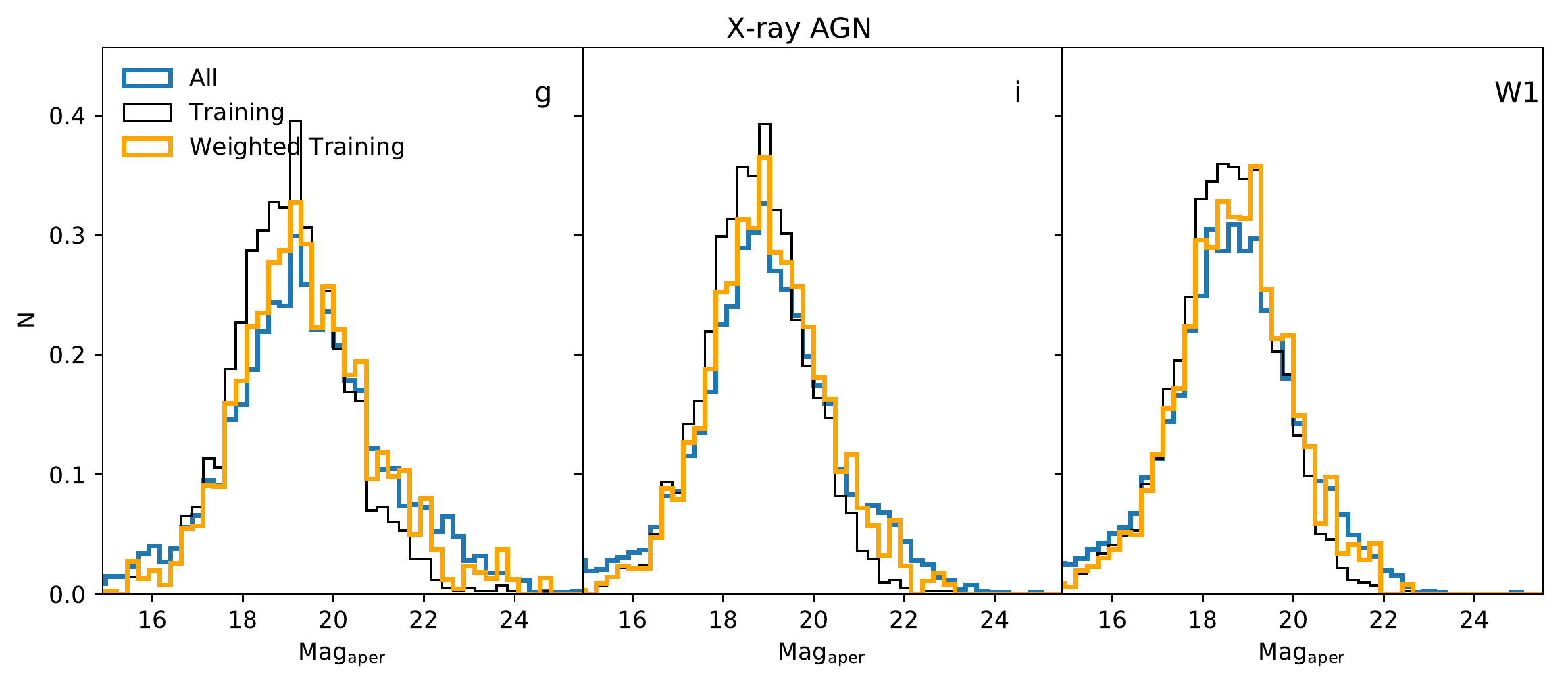}
  \includegraphics[width=1\columnwidth]{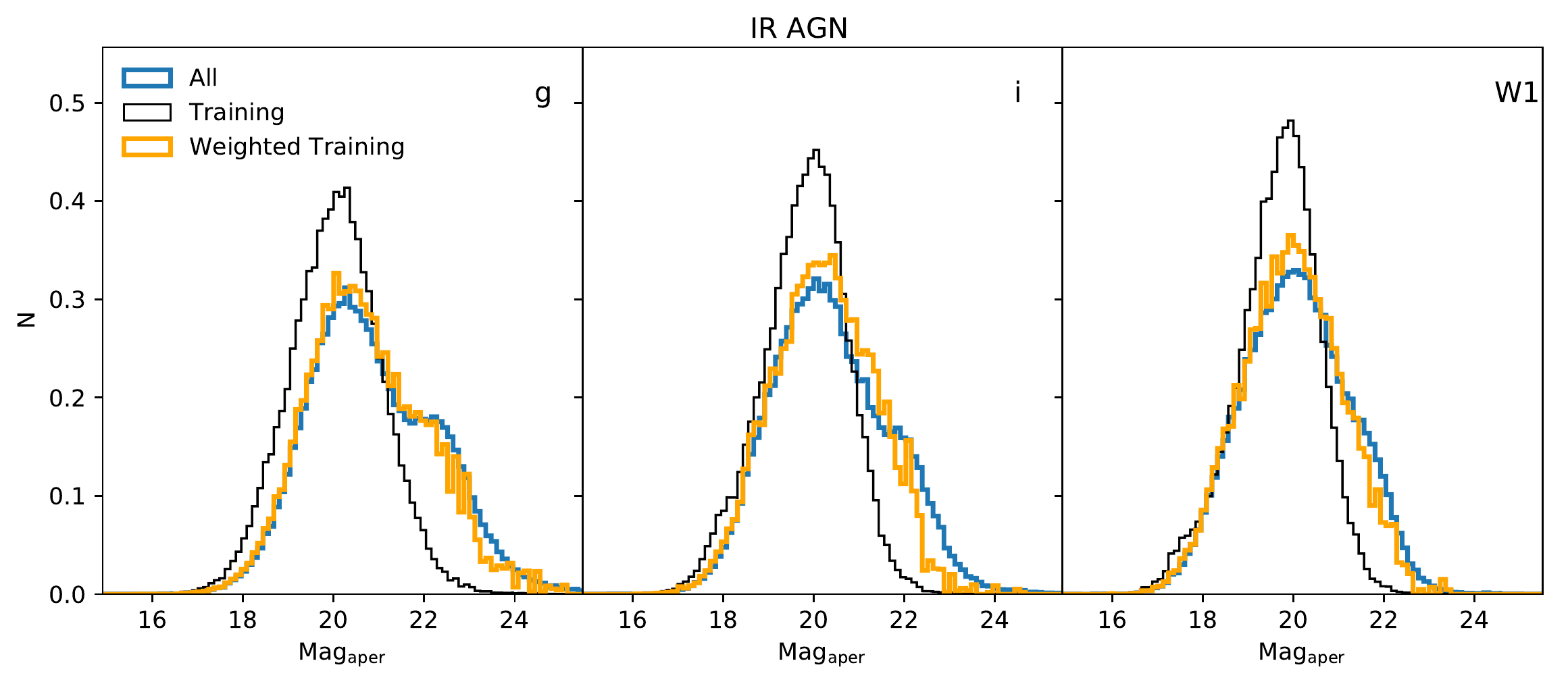}
  \caption{Illustration of the colour-magnitude based weighting scheme applied to the AGN training subsets employed in this work. The thick blue line shows the magnitude distributions for the full photometric sample while the thin black and thick gold lines show the training sample before and after weighting. For each magnitude distribution, the corresponding photometric band is labelled in the upper right corner of the panel. Compared to the non-AGN population, the overall weighting required is relatively small.}
  \label{fig:weights_agn}
\end{figure}

As with the galaxy \textsc{GPz} estimates, we calculate colour and magnitude dependent weights that are incorporated during training through cost-sensitive learning.
When calculating the training sample weights for the AGN subsets, we make use of the $g-i$ and $i-W1$ colours combined with the $i$-band magnitude.
The results of the training sample weights for the AGN subsets are presented in Fig.~\ref{fig:weights_agn}.
Compared to the `normal' optical galaxy population, the training sample for the AGN selected subsets are significantly less biased.
Nevertheless, we find that our weighting scheme still helps to bring the training sample into much closer agreement with the full photometric sample.

\begin{figure*}
\centering
  \includegraphics[width=0.25\paperwidth]{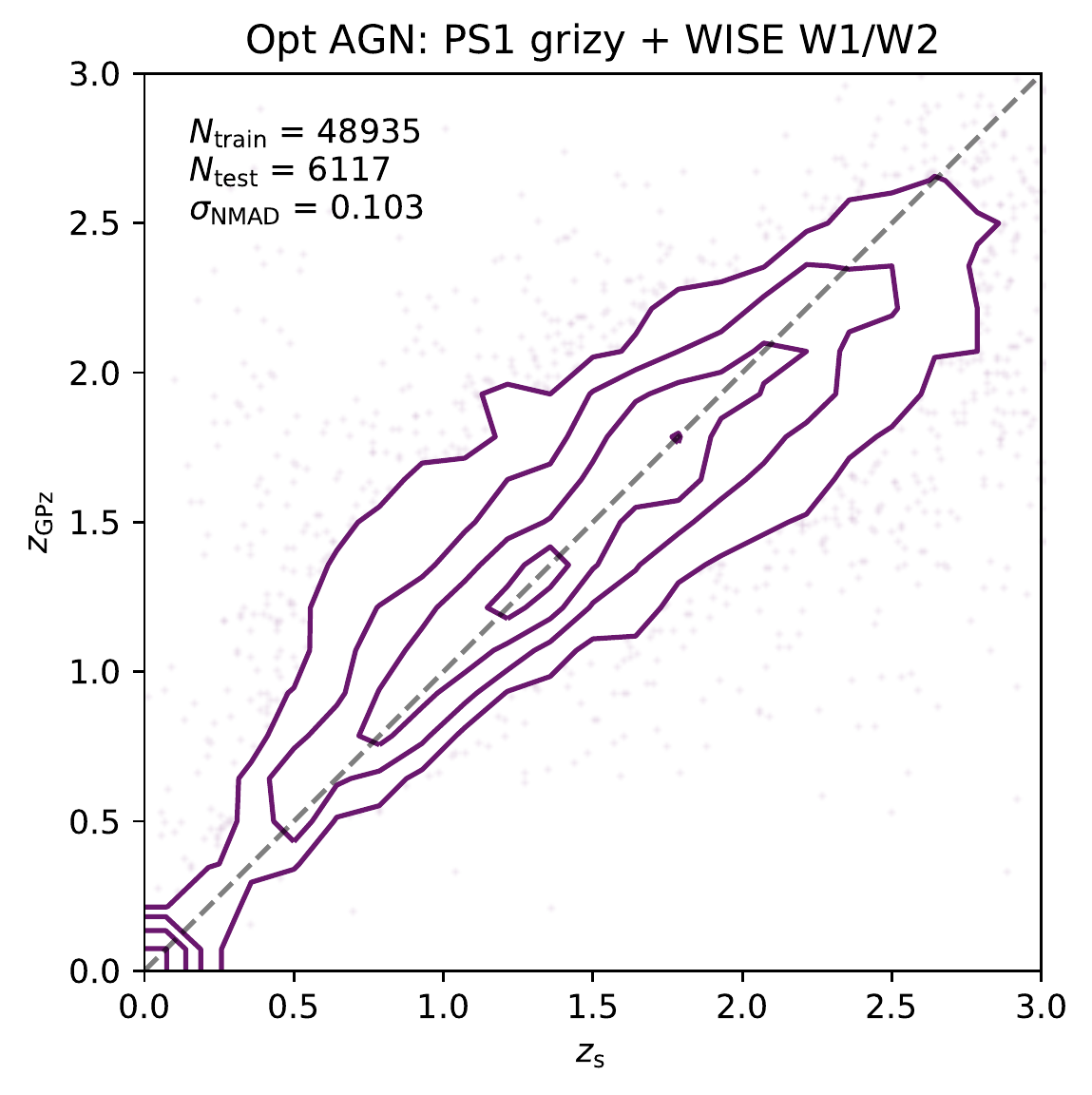}
  \includegraphics[width=0.25\paperwidth]{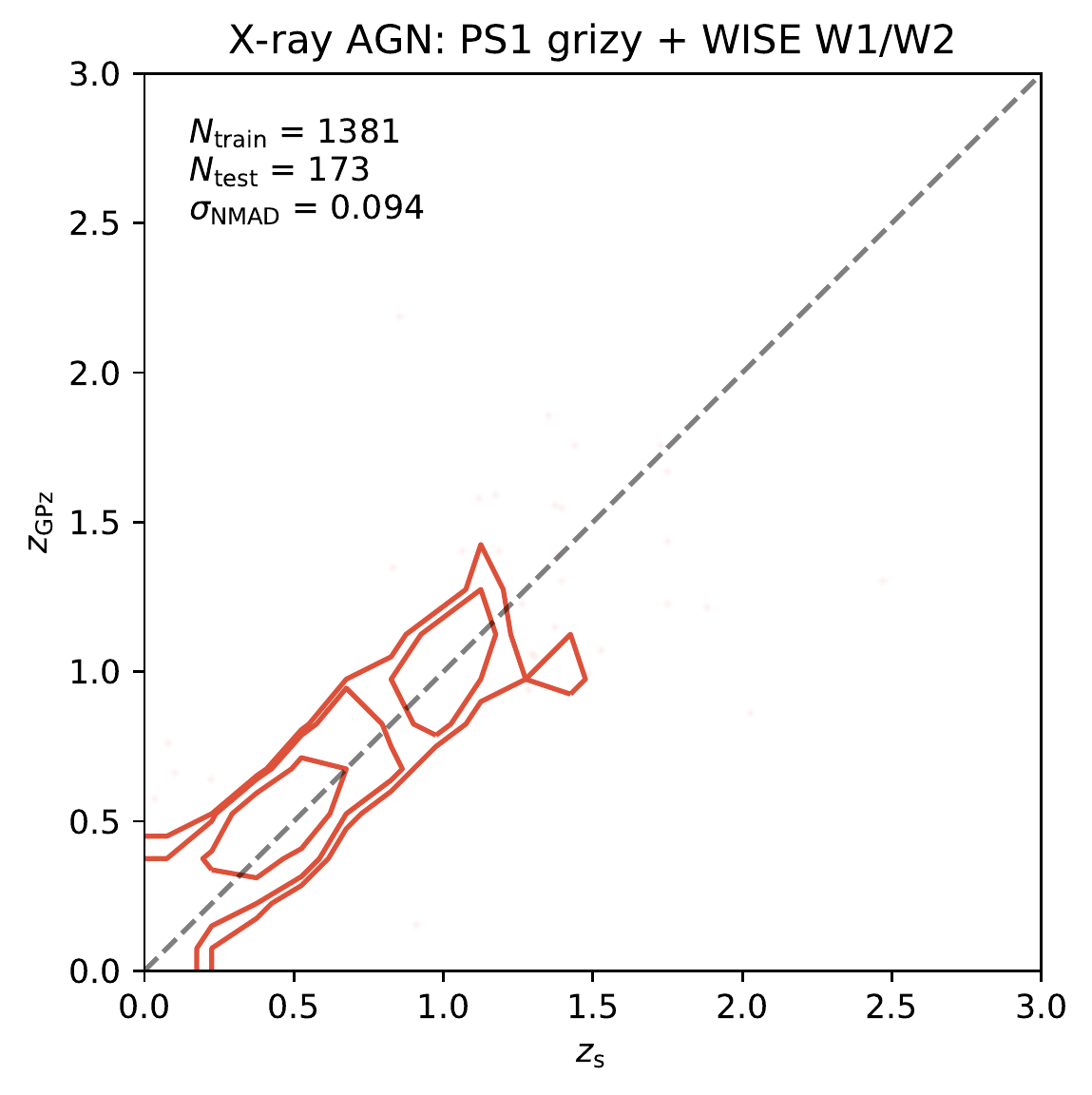}
  \includegraphics[width=0.25\paperwidth]{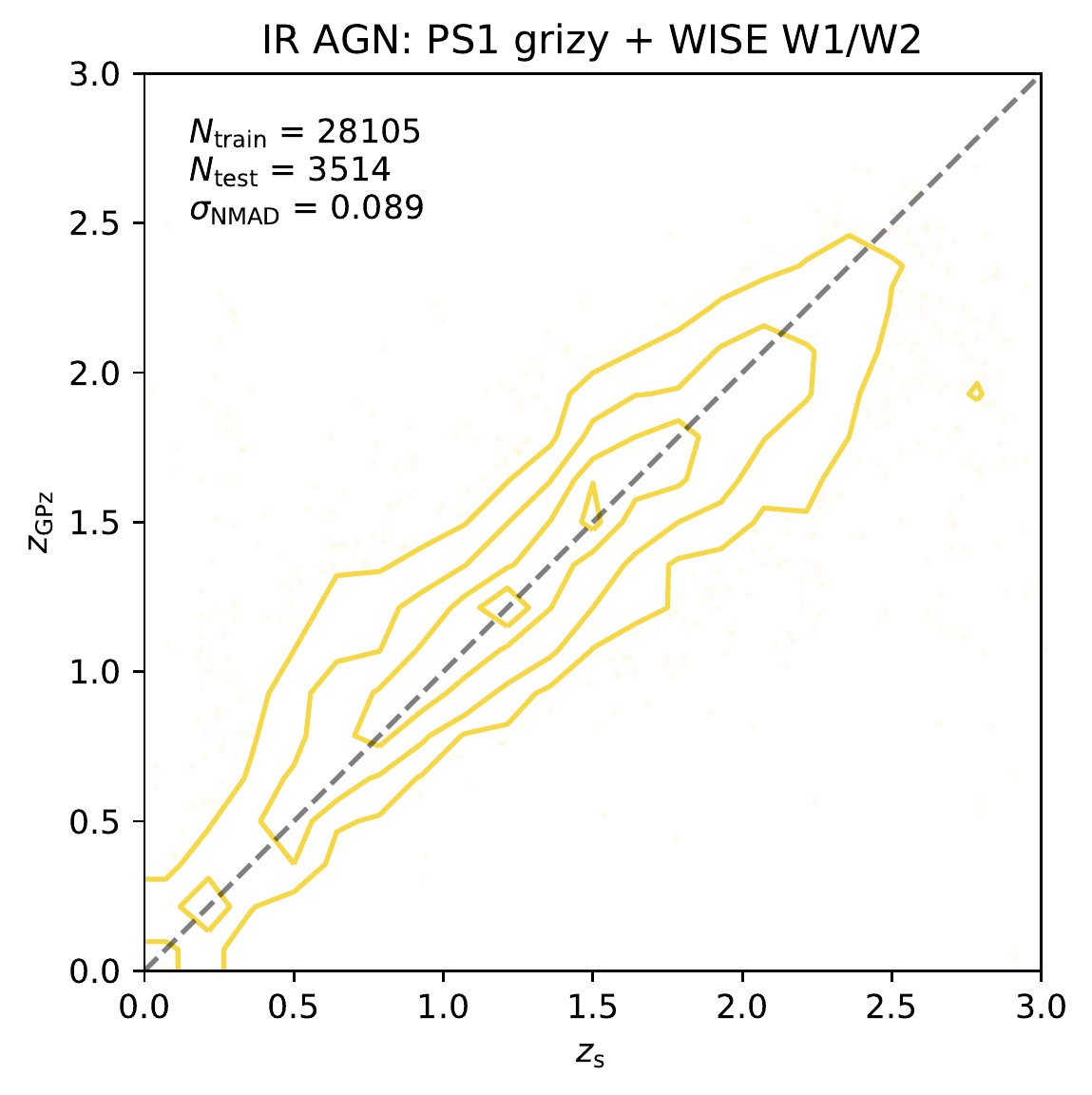}
  \caption{Distribution of \textsc{GPz} photometric redshift estimates vs spectroscopic redshift for the test sample (not included in training in any way) for the three AGN subsamples; optically identified quasars (left), X-ray selected AGN (centre) and WISE infrared selected AGN (right). The number of training sources used ($N_{\textup{train}}$), the number of test sources plotted ($N_{\textup{test}}$) and the corresponding robust scatter for the test sample ($\sigma_{\textup{NMAD}}$) are shown in the the upper left corner of each panel. The plotted contours are linearly spaced in source density.}
  \label{fig:gpz_agn_speczphotz}
\end{figure*}

In Fig.~\ref{fig:gpz_agn_speczphotz} we show the resulting photometric vs spectroscopic redshift distributions for each of the AGN subset-specific \textsc{GPz} estimates.
For both the optical and infrared selected AGN samples (for which there is extensive overlap within the training sample; Fig.~\ref{fig:agn_venn_spec}), the spectroscopic training sample extends out to high redshift.
In line with expectations for the AGN population selected by these criteria \citepalias{Duncan:2017wu}, the robust scatter with respect to the spectroscopic sample is worse than for the galaxy population.
However, the overall performance is very good for the AGN population and competitive with studies in the literature with similar or better datasets \citep[e.g.][]{Richards:2001ct,Brodwin:2006dp,2012MNRAS.424.2876M,Chung:2014it}.

Due to the relatively small number of spectroscopic sources at the very highest redshifts ($z \gtrsim 2.5$), the panels in Fig.~\ref{fig:gpz_agn_speczphotz} do not clearly illustrate the poorer performance of the \textsc{GPz} estimates in this regime; where the estimates become increasingly biased (towards spuriously low photo-$z$s).
This is a well known limitation of empirical photo-$z$ methods due to the sparser training samples available at these highest redshifts and is discussed in greater detail in \citetalias{Duncan:2017ul}. 
However, it is at high redshift where the strong optical features are expected to enable good photo-$z$ estimates from template fitting methods - hence the motivation for the hybrid methodology employed in this work.


\subsection{Template-fitting Estimates}\label{sec:method-temp}
The template-fitting photometric redshifts are estimated following the method outlined in \citetalias{Duncan:2017wu}.
For the purposes of this paper we present a brief summary of the method and outline key changes in its application to the HETDEX dataset.

We calculate photometric redshifts using three different galaxy template sets from the literature that are either commonly used in photometric redshift estimates within the literature and/or designed to cover the broad range of SEDs observed in local galaxies.

The three template sets used in this analysis are as follows:
\begin{enumerate}
	\item \citet{Brammer:2008gn} default \textsc{eazy} reduced template set (`EAZY') - The first set used are the updated optimised \textsc{eazy} template set  that includes galaxy templates with stellar emission only.
	
	\item \citet{Salvato:2008ef} `XMM-COSMOS' templates - Our second set of templates is that presented by \citet{Salvato:2008ef,Salvato:2011dq}, including 30 SEDs that cover a wide range of galaxy spectral types in addition to both AGN and QSO templates.
The XMM-COSMOS templates include both dust continuum and PAH features as well as power-law continuum emission for the appropriate AGN templates.

	\item \citet{Brown:2014jd} Atlas of Galaxy SEDs (`Atlas') - The large atlas of 129 galaxy SED templates presented in \citet[][referred to as `Atlas' hereafter]{Brown:2014jd}.
Designed to sample the full colour space of nearby galaxies, the `Atlas' templates cover a broad range of galaxy spectral types including ellipticals, spirals and luminous infrared galaxies (both starburst and AGN).
\end{enumerate}

As the \textsc{eazy} templates include only stellar emission we fit only to the PS1 optical and WISE W1/W2 bands; excluding the WISE mid-IR photometry that may be contaminated by sources of non-stellar radiation in low redshift sources.
When fitting the XMM-COSMOS and Atlas templates, the WISE W3 ($12\mu$m) band is allowed in the redshift fitting.
Due to the extensive problems with source confusion, W4 ($12\mu$m) is not included in any of the fits.
We note however that given the sensitivity limits of the various bands, WISE W3 and W4 detections exist only for a very small subset of the complete AllWISE catalog.

We include the additional rest-frame wavelength dependent flux errors using the \textsc{eazy} template error function \citep[see][]{Brammer:2008gn} for all fits (ranging from $<5\%$ at rest-frame optical wavelengths to $>15\%$ at rest-frame UV and near-IR).
During the template fitting, zeropoint offsets are calculated based on the full spectroscopic training sample for both the AGN and non-AGN population. 
For all three template sets, offsets to the PS1 and W1/W2 filters are in the range of $\sim1-3\%$ (with the exception of an offset of $+5.3\%$ for W2 for the XMM-COSMOS set). 
Offsets to the W3 band are $+3.6\%$  and $+30\%$ for the Atlas and XMM-COSMOS template sets, respectively. 

Finally, due to the reduced number of optical filters available for template-fitting compared to the deep fields on which this method was originally tested, a magnitude dependent redshift prior \emph{is} included within the individual template estimates.
The magnitude-dependent prior is particularly beneficial for this dataset due to the lack of $u$-band photometry probing rest-frame features below the 4000\AA~ break for the low-redshift population resulting in increased confusion between the 4000\AA/Balmer and Lyman break features.
For optically bright low-redshift galaxies, the magnitude prior is able to rule out implausible redshift solutions (i.e. $z > 1$).
Our magnitude-dependent redshift prior functions for the separate AGN and galaxy subsets are calculated for the PS1 $i$-band following the procedure in Section~5.1.1 of \citetalias{Duncan:2017wu}.

\begin{figure*}
\centering
\includegraphics[width=0.28\paperwidth]{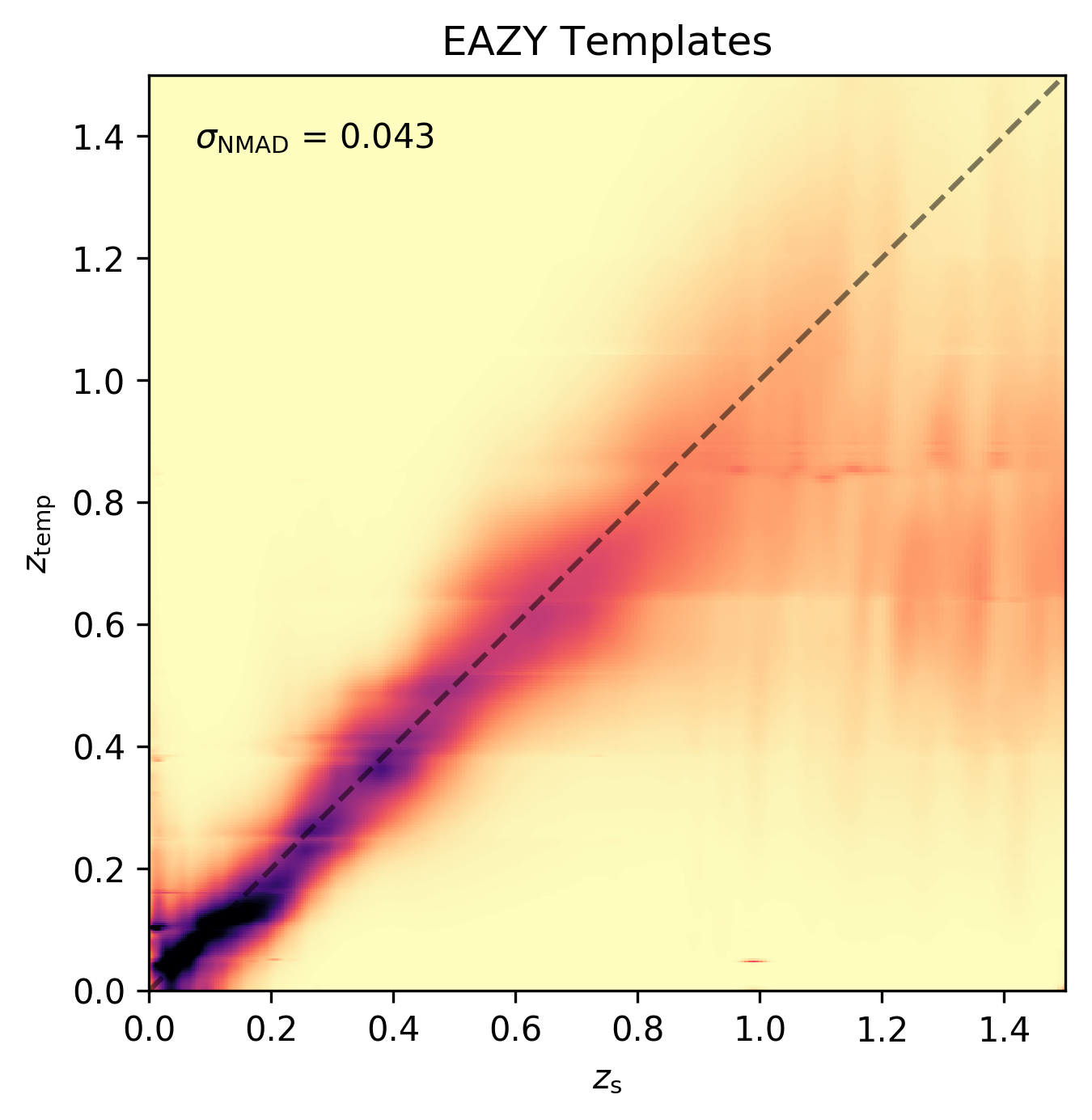}
\includegraphics[width=0.28\paperwidth]{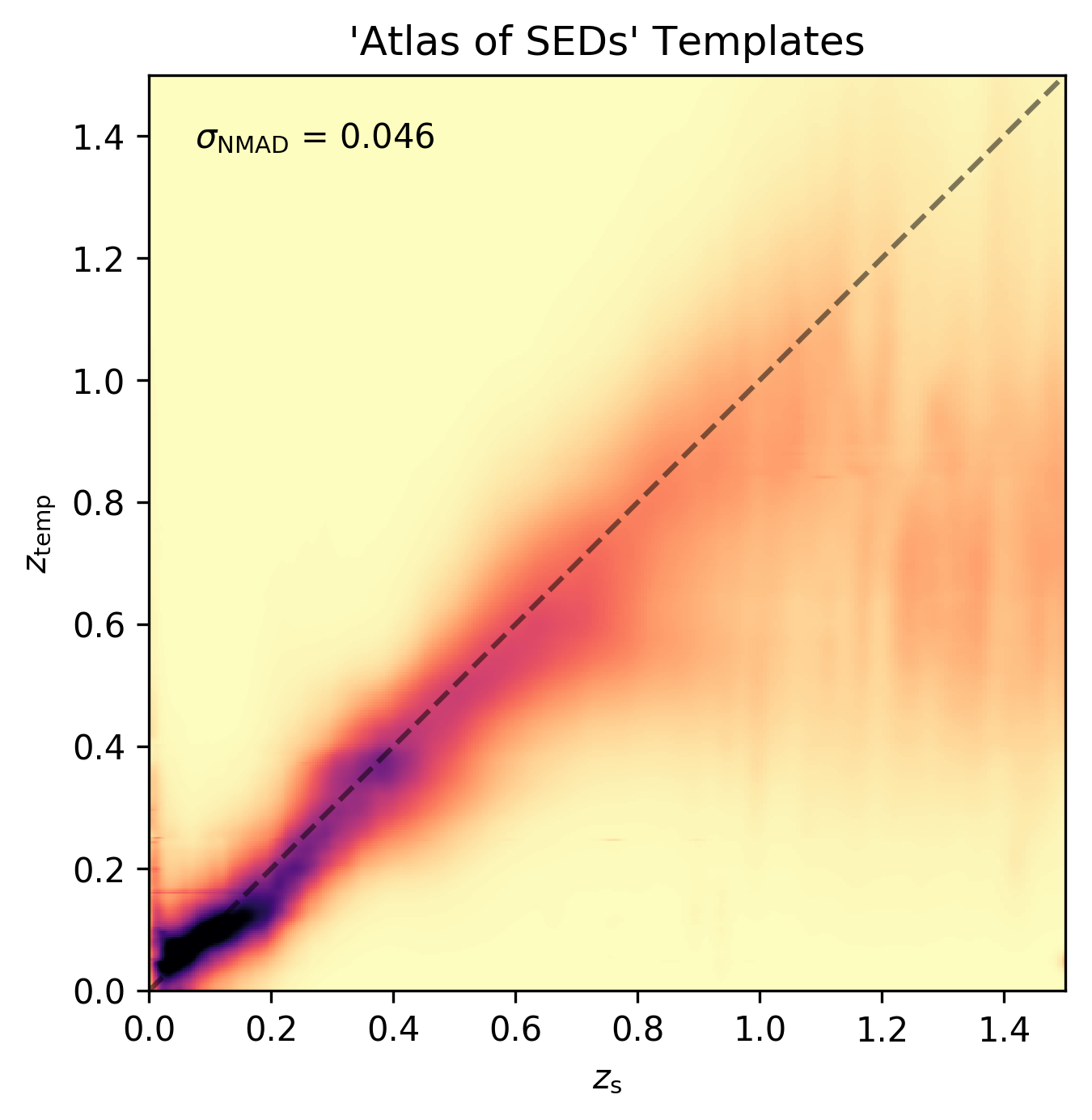}
\includegraphics[width=0.28\paperwidth]{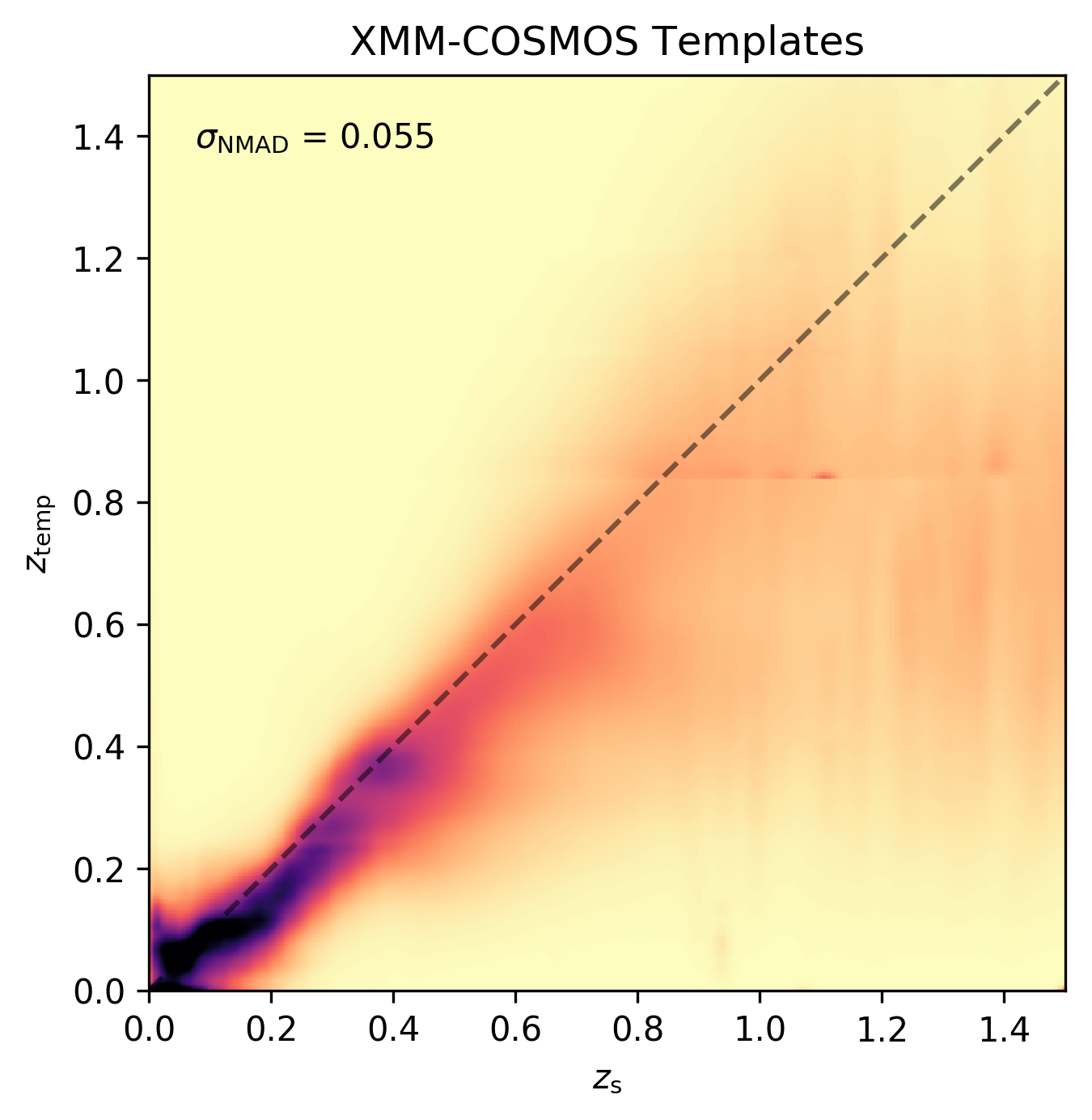}
  \caption{Stacked template-fitting posterior redshift predictions for the host-dominated galaxy population for each of the template sets used. To improve the visual clarity at higher redshifts where there are few sources within a given spectroscopic redshift bin, the distributions have been smoothed along the x-axis.}
  \label{fig:template_photzspecz_gal}
\end{figure*}

In Fig~\ref{fig:template_photzspecz_gal} we present a qualitative illustration of the three template-based photo-$z$ estimates for the non-AGN population.
As the template-fitting method results in a full redshift posterior prediction rather than a single Gaussian prediction, Fig~\ref{fig:template_photzspecz_gal} shows the stacked redshift posteriors in bins of spectroscopic redshift.
We see that broadly speaking the template photo-$z$ performance is better for the \textsc{Eazy} default template library than for other two libraries.
All three estimates however are worse than the empirical \textsc{GPz} estimates for the same subset.

\begin{figure}
\centering
\includegraphics[width=0.9\columnwidth]{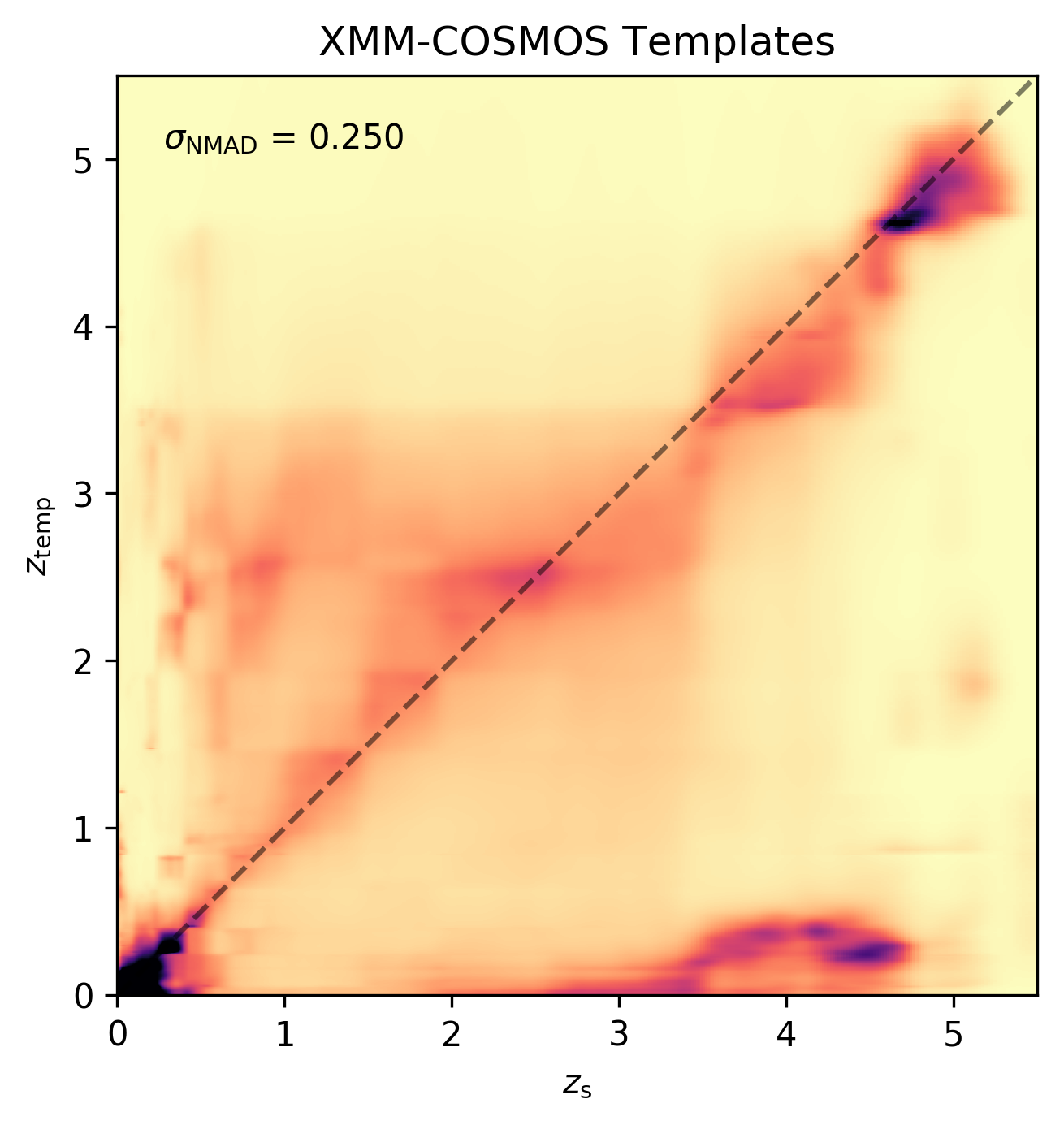}
  \caption{Stacked template-fitting posterior redshift predictions for the combined AGN selected population (IR, X-ray or optically selected). To improve the visual clarity at higher redshifts where there are few sources within a given spectroscopic redshift bin, the distributions have been smoothed along the x-axis.}
  \label{fig:template_photzspecz_agn}
\end{figure}

For the subset of sources that satisfy one or more of the AGN selection criteria, the performance of the template estimates is even poorer - so much so that the \textsc{Eazy} and `Atlas' template estimates are un-useable.
The reason for this poor performance can be attributed to the nature of the AGN spectroscopic training sample and the dominance of optically bright quasars within it; a spectral type that is only included in the XMM-COSMOS library.
Although the other template sets may still provide useable estimates for non-QSO sources \citepalias[as seen in][]{Duncan:2017wu}, we conservatively choose to incorporate only the XMM-COSMOS photo-$z$ estimates (see Fig.~\ref{fig:template_photzspecz_agn}) within the subsequent Bayesian combination analysis for AGN sources.

\subsection{Hierarchical Bayesian combination}\label{sec:hbcombination}
To produce the final consensus redshift prediction for a given source, we use the Hierarchical Bayesian (HB) combination method outlined in \citetalias{Duncan:2017wu} \citep[based on the method presented in][]{Dahlen:2013eu} and subsequently extended to hybrid \textsc{GPz} + template estimates in \citetalias{Duncan:2017ul}.
In summary, hierarchical Bayesian combination produces a consensus redshift prediction, $P(z)$, from a set of $n$ individual predictions while accounting that for the possibility that any individual \emph{measured} redshift posterior distribution $P_{m}(z)_{i}$ is incorrect.
The possibility that an individual $P(z)$ is incorrect is introduced as a nuisance parameter, $f_{\text{bad}}$, and in the case where a measurement is incorrect, a prior on the redshift distribution is assumed.
The final consensus redshift is then obtained by marginalising over the nuisance parameter.
The plausible range of $f_{\text{bad}}$ and the relative covariance between the different estimates, $\beta$ ($<=n$), are hyper-parameters that can be optimised using training data such that the posterior redshift distributions more accurately represent the redshift uncertainties.

During the HB procedure, \textsc{GPz} estimates are converted to the same redshift grid as used during the template fitting procedure by evaluating normal distributions based on \textsc{GPz} predicted centre $z_{\textup{GPz}}$ and corrected variance estimate.
As in \citetalias{Duncan:2017ul}, if a source does not have a photo-$z$ estimate for a given \textsc{GPz} estimator (either through not satisfying the selection criteria for a given subset or lack of observations in a required band) it is assumed to have a flat redshift posterior.
\textsc{GPz} therefore contributes no additional information to the consensus HB estimates for these sources. 

For the application in this work, based on the outlier fractions in trial runs of the consensus redshift estimates we assume $0 \leq f_{\text{bad}} \leq 0.05$ and $0 \leq f_{\text{bad}} \leq 0.2$ for the galaxy and AGN subsets respectively.
Using the spectroscopic training sample the optimum choices for the hyperparameter, $\beta$, were found to be $\beta = 4.2$ for galaxies and $\beta = 1$ for AGN.
After testing the Bayesian combination with a flat, volume-element and magnitude-dependent prior assumption for `bad' estimates (see \citetalias{Duncan:2017wu}), a flat prior on the redshift distribution for the HETDEX sample was found to produce better results.

\subsection{Calibration of Photo-$z$ uncertainty}
Correctly calibrating the uncertainties on photo-$z$ \citep{Dahlen:2013eu,2016MNRAS.457.4005W} is crucial, both scientifically and for the Bayesian combination procedure.
To quantify the over- or under-confidence of our photometric redshift estimates, we follow the method outlined in Section~3.3.1 of \citetalias{Duncan:2017ul} \citep[and originally proposed in ][]{2016MNRAS.457.4005W} and calculate the distribution of threshold credible intervals, $c$, where the spectroscopic redshift intersects the redshift posterior.
For a set of redshift posterior predictions which perfectly represent the redshift uncertainty (e.g. 10\% of galaxies have the true redshift within the 10\% credible interval, 20\% within their 20\% credible interval, etc.), the expected distribution of $c$ values should be constant between 0 and 1.
The cumulative distribution, $\hat{F}(c)$, should therefore follow a straight 1:1 relation, i.e. a Q-Q plot.
Curves which fall below this expected 1:1 relation therefore indicate that there is overconfidence in the photometric redshift errors; the $P(z)$s are too sharp.

\subsubsection{Uncertainty calibration for \textsc{GPz} estimates}\label{sec:gpz_error_calibration}
As in \citetalias{Duncan:2017ul} we calculate the threshold credible interval for the \textsc{GPz} predictions analytically as:
\begin{equation}
c_{i} = \Phi(n_{i}) - \Phi(-n_{i}) = \textup{erf} \left (\frac{n_{i}}{\sqrt{2}} \right),
\end{equation}\label{eq:gpz_ci_calc}
\noindent where $\Phi(n_{i})$ is the normal cumulative distribution function and $n_{i}$ can be simply calculated as $| z_{i,\textup{spec}} - z_{i,\textup{phot}} | / \sigma_{i}$.
We then scale the uncertainties, $\sigma_{i}$, as a function of magnitude, $m_{i}$, such that 
\begin{equation}
	\sigma_{\textup{new},i} = \sigma_{\textup{old},i} \times \alpha(m_{i}).
\end{equation}\label{eq:gpz_err_scale}
\noindent The magnitude dependence assumes the relation
\begin{equation}
	\alpha(m) = \begin{cases}
	 \alpha_{\eta} & m \leq m_{\eta}\\
	 \alpha_{\eta} + \kappa \times(m-m_{\eta}) & m > m_{\eta}.
	\end{cases}
\end{equation}\label{eq:smoothing_2}
\noindent where $\alpha(m)$ is a constant value, $\eta$, below some characteristic apparent magnitude, $m_{\eta}$, and follows a simple linear relation above this magnitude \citep{2009ApJ...690.1236I}.
We use the PS1 $i$-band optical magnitude for calculating the magnitude dependence of the error scaling and assume a characteristic magnitude of $i = 16$.
The parameters $\eta$ and $\kappa$ are then fit using the \textsc{emcee} Markov Chain Monte Carlo fitting tool \citep[MCMC;][]{ForemanMackey:2013io} to minimise the Euclidean distance between the measured and ideal distributions.

After calibrating using the training and validation subsets, we find that the calibrated uncertainties for the test sample for each subset (optical/X-ray/AGN/galaxies) are significantly improved and lie close to the desired 1:1 relation. 
However, even after calibration we find that the very wings of the posterior distribution are slightly under-estimated.
At the very faintest magnitudes the uncertainties become significantly overestimated (we are under-confident) for most subsets, but particularly the galaxy subsets.
As the individual \textsc{GPz} estimates represent an intermediate step we do not include illustration of the individual uncertainty calibrations here.
In Section~\ref{sec:pz_accuracy} we will present the uncertainty distributions for the final calibrated consensus redshift posteriors.

\subsubsection{Uncertainty calibration for the template estimates}\label{sec:template_error_calibration}
Calibration of the template uncertainties is performed in a similar manner, using a modified version of the procedure outlined in \citetalias{Duncan:2017wu}.
Due to the inclusion of the magnitude dependent redshift prior in this work ($P(z|m_{i})$), we define the optimised posterior redshift for a given source, $i$,  as
\begin{equation}
	P(z)_{\textup{new}, i} \propto P(z)_{\textup{old}, i}^{1/\alpha(m_{i})} \times P(z|m_{i}),
\end{equation}\label{eq:smoothing_1}
where $\alpha(m)$ follows the relation described in Equation~\ref{eq:smoothing_2} and the parameters $\eta$ and $\kappa$ are optimised in the same way as described above.

Due to the prohibitively long computation time required to use the full spectroscopic sample, we use only a subset for the purposes of template error calibration.
For both the AGN and galaxy samples separately, a subset of each training sample is created by randomly selecting up to 3000 sources in each of 6 magnitude bins over the range covered by the spectroscopic redshift subset ($16 < i < 22$).
Calibration of the uncertainties is then done on $2/3$ of this subsample, with the other $1/3$ retained for testing.

Tests of the error calibration using a range of smaller sub-samples suggest that the accuracy of the uncertainties after calibration is likely not affected by the sample size.
Specifically, we find that accuracy of the uncertainties, as quantified by the Euclidean distance between the measured $\hat{F}(c)$ distribution after calibration and the desired 1:1 relation, is not a strong function of the size of the sample for subsets of between 100 and 750 sources per magnitude bin.
 We note however that these tests (and the final calibrated estimates) are still limited by how representative the available spectroscopic training and test sample is of the full photometric sample -- with this bias likely representing the major systematic limitation on the accuracy of the uncertainties.

\subsection{Accuracy of the photo-$z$ uncertainties}\label{sec:pz_accuracy}
After calibration of the individual input estimates, the final stage of the uncertainty calibration comes as part of the tuning of the hierarchical Bayesian combination hyper-parameters, specifically $\beta$ (see Section~\ref{sec:hbcombination}).

\begin{figure*}
\centering
  \includegraphics[width=0.9\textwidth]{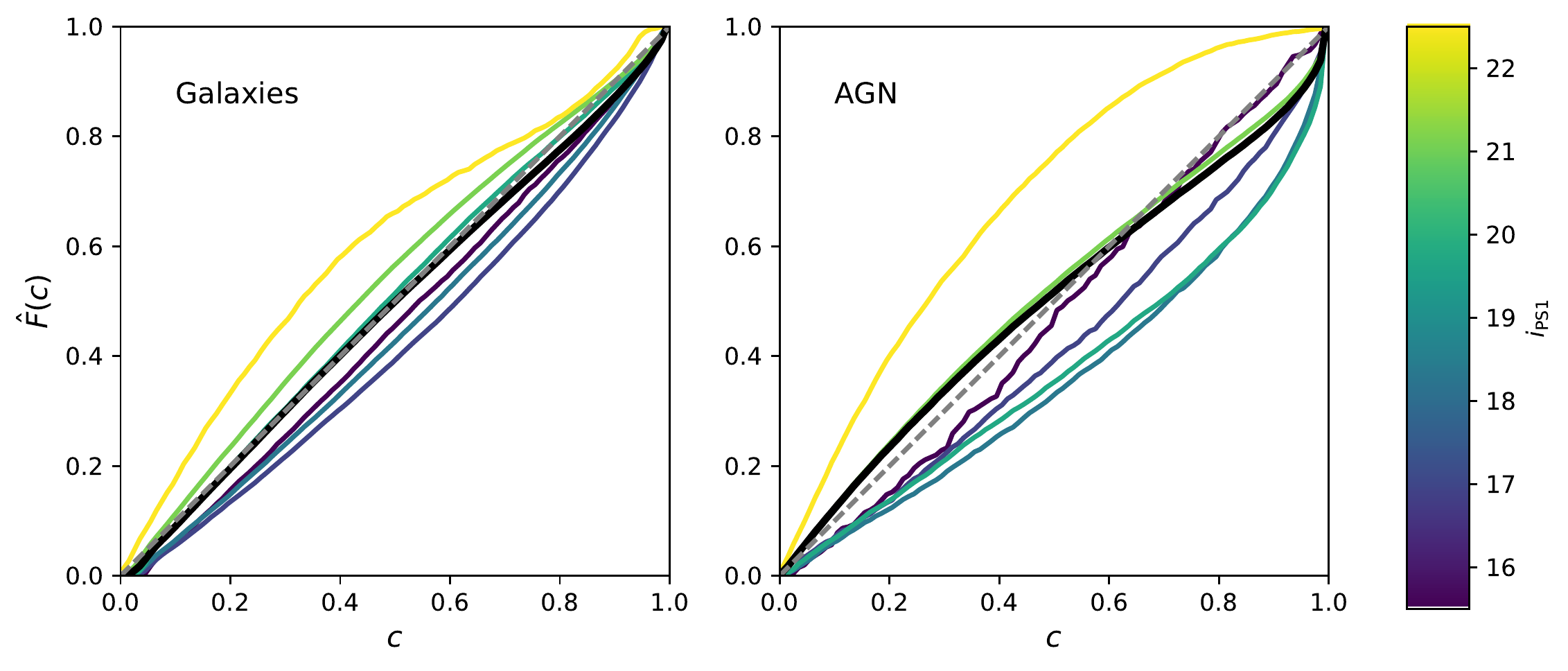}
  \caption{Q-Q ($\hat{F}(c)$, see text in Section~\ref{sec:method-gpz}) plots for  the final calibrated consensus redshift predictions for the galaxy population (left) and the optical/infrared and X-ray selected AGN population (right). Coloured lines represent the distributions in bins of apparent optical magnitude while the thick black line corresponds to the complete spectroscopic training sample. Lines that fall above the 1:1 relation illustrate under-confidence in the photo-$z$ uncertainties (uncertainties overestimated) while lines under illustrate over-confidence (uncertainties underestimated).}
  \label{fig:pz_accuracy}
\end{figure*}

In Fig.~\ref{fig:pz_accuracy} we illustrate the accuracy of the final calibrated redshift posteriors for the AGN and galaxy subsets.
Shown in both plots are the cumulative distribution ($\hat{F}(c)$) of threshold credible intervals, $c$, both for the full spectroscopic sample (thick black lines) and within bins of apparent magnitude (coloured lines).

For both subsets, the uncertainties for the whole spectroscopic sample are well calibrated, lying close to the desired 1:1 relation.
However, we can see that there are still some residual trends as a function of apparent magnitude.
For the galaxy population, the magnitude trend is relatively small with all but the very faintest magnitudes close to ideal trend.
For the AGN population this trend is more stark, with a rapid evolution as a function of $i_{\textup{PS1}}$ magnitude leading to significant under-confidence in the uncertainties for the faint sources.

\section{Photometric Redshift Properties}\label{sec:results}
\begin{figure}
\centering
  \includegraphics[width=0.9\columnwidth]{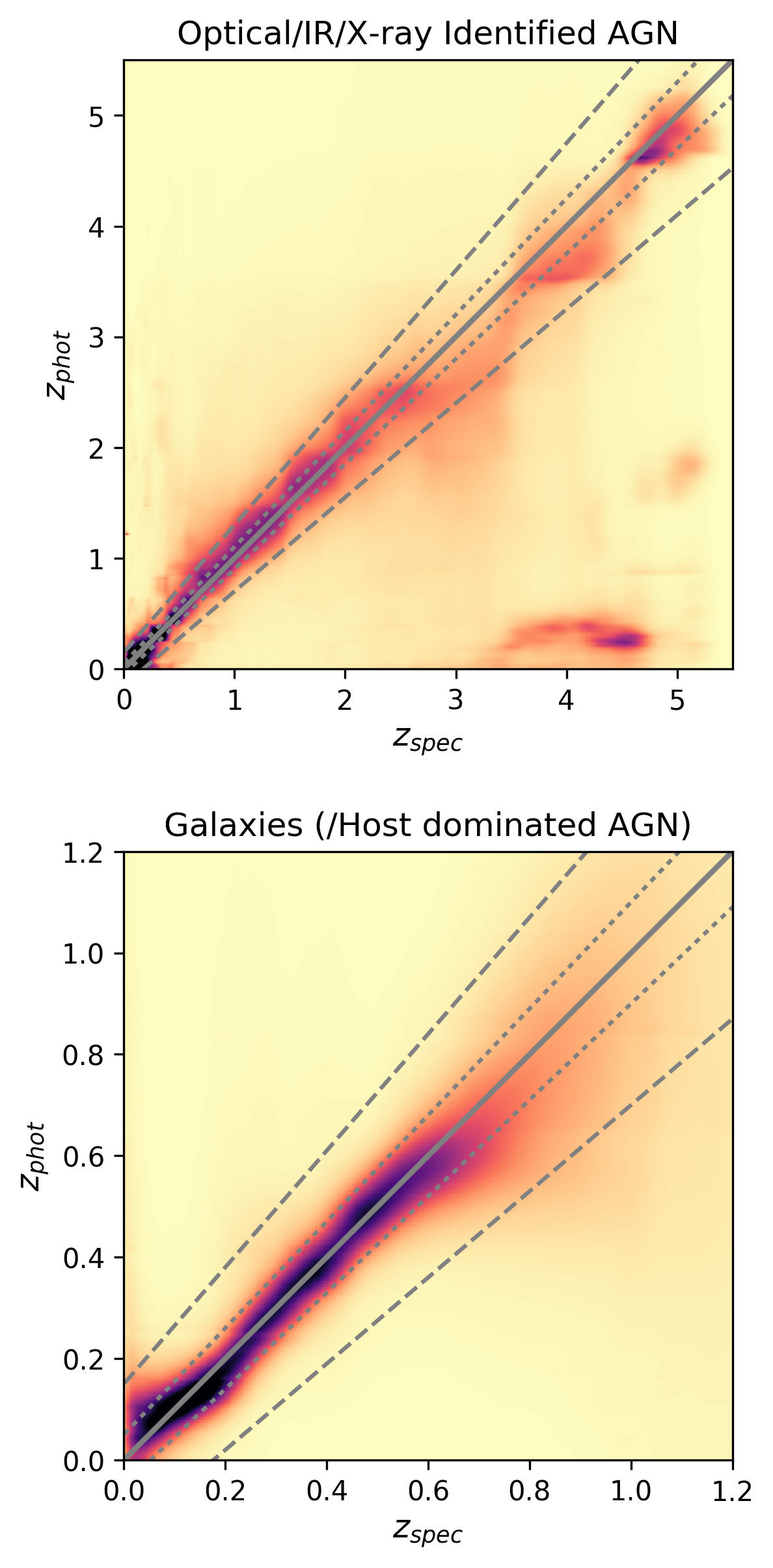}
  \caption{Stacked probability distributions for the combined AGN selected population (top; IR, X-ray or optically selected) and the normal galaxy (or host-dominated) population as a function of spectroscopic redshift for the consensus HB photo-$z$ estimate. To improve the visual clarity at higher redshifts where there are few sources within a given spectroscopic redshift bin, the distributions have been smoothed along the x-axis. The solid grey line corresponds to the desired 1:1 relation while the dotted and dashed lines correspond to $\pm 0.05 \times (1+z_{spec})$ and $\pm 0.15 \times (1+z_{spec})$ respectively.}
  \label{fig:hb_pz_stack}
\end{figure}

After the error calibration for all input estimates and the tuning of the Bayesian combination hyper-parameters, we calculate consensus estimates for the entire photometric catalog.
In Fig.~\ref{fig:hb_pz_stack} we present a qualitative illustration of the final consensus redshifts for the spectroscopic training sample.
We show the stacked redshift posteriors as a function of spectroscopic redshift for both the multi-wavelength AGN subset (top) and for the remaining galaxy population (bottom).
For the AGN subset, the plot clearly shows the significant improvements offered by the hybrid methodology, with posteriors for the $z_{\textup{spec}} \lesssim 2.5$ matching those of the \textsc{GPz} estimates seen in Fig~\ref{fig:gpz_agn_speczphotz}.
At higher redshifts where the \textsc{GPz} estimates become significantly biased, the hybrid estimates are able to key into the strong Lyman-break feature and provide better estimates.

We can see, however, that the redshift estimates are not perfect.
At $3.5 \lesssim z_{\textup{spec}} \lesssim 4.5$ there is a cluster of sources for which there is a catastrophic failure in the redshift estimates - with posterior predictions of $z\sim 0.3$.
These sources represent a minority of the spectroscopic sample at high redshift: of the sources that have $z_{\textrm{spec}} > 3.5$ (1019), we find that only $\approx 10\%$ (116) of sources are truly catastrophic outliers with neither primary nor secondary redshift solutions within $z_{\textup{spec}} \pm 0.3\times(1+z_{\textup{spec}})$.
Of these catastrophic failures, 73 have $z_{1,\textup{median}} < 1$ and contribute to the cluster seen at $z\sim 0.3$, representing $7.2\%$ of the $z_{\textrm{spec}} > 3.5$ sample .

Investigating the properties of these outliers with respect to the sources that have accurate predictions reveals no clear single origin for the poor predictions.
Their overall colour distribution does not differ significantly from the sources that are well fitted.
However, we find that these sources are disproportionately brighter than the majority of the spectroscopic QSOs at these redshifts -- with apparent magnitudes of $i < 20$.

Between $2.5 \lesssim z_{\textup{spec}} \lesssim 4$ the redshifts that \emph{are} well fit still become noticeably more biased and have large uncertainty (as illustrated by the broad $z_{\textup{phot}}$ distribution).
The increased uncertainty in this redshift range likely results from the lack of $u$-band photometry in the PanSTARRS data - whereby the Lyman break is not probed until $z\sim 4$.
The relative sparsity of training sources available at the redshift mean that \textsc{GPz} is not able to compensate in this regime.
Significantly greater numbers of training sources in this regime may allow future implementations to overcome this by improving the \textsc{GPz} estimates.
Alternatively, additional $u$-band photometry could be included within the dataset to improve the precision of both methods.
The strict non-detection in $u$-band for $z > 3$ sources may also help to break any colour degeneracies causing the catastrophic failures at $z_{1,\textup{median}} \sim 0.3$.

For the `Galaxy' sample (Fig~\ref{fig:hb_pz_stack} lower panel) we see that the consensus redshift estimates are excellent over the redshift range $0.1 \lesssim z_{\textup{spec}} \lesssim 0.8$, with very low scatter and very little bias.
Beyond $z_{\textup{spec}} \lesssim 0.8$ the posteriors become increasingly broad.
As illustrated by Fig.~\ref{fig:specz_hist_gal}, this transition redshift represents the limits of the spectroscopic training (and test) sample and also potentially the limits of the optical photometry itself.
In the following section we explore these limitations in more detail with a more quantitative analysis of the photo-$z$ estimates.

\subsection{Overall photo-$z$ statistics}
While the $z_{\textup{spec}}$ vs $z_{\textup{phot}}$ plots are helpful in qualitatively assessing the quality of the $z_{\textup{phot}}$ estimates and identifying any major problems, a more quantitative analysis is required to enable both comparison with other estimates (if available) and for the user to judge reasonable selection criteria for their science samples.

It is common within the literature to judge the quality of photo-$z$s by comparing a single valued `best' estimate for the photo-$z$.
Reducing the full posterior redshift prediction to a single value has inherent problems because it can potentially present a biased view of that posterior prediction and is effectively throwing away information.

Nevertheless, to enable the comparison we must first choose a way to represent the redshift posteriors in a format suitable for catalogs and single-value based quality statistics.
Common practice is to take either the maximum a posteriori value for the redshift prediction, the median of the redshift posterior or the expected value of the posterior (these can differ significantly in the case of skewed posteriors or secondary redshift solutions).

In the catalogs and the subsequent analysis, we take an approach motivated by the discussion of \citet{2016MNRAS.457.4005W} and aimed at providing an accurate representation of the redshift posteriors.
For each calibrated redshift prediction, we first calculate the 80\% highest probability density (HPD) credible interval (CI) by starting at the redshift peak probability and lowering a threshold until 80\% of the integrated probability is included.
Next, we identify the primary peak (and secondary peak if present) by identifying the points where the $P(z)$ cross this threshold.
For each peak, we then calculate the median redshift within the boundaries of the 80\% HPD CI to produce our point-estimate of the photo-$z$ (hereafter $z_{1,\textup{median}}$ or $z_{2,\textup{median}}$). 
As a measure of the redshift uncertainty, in the catalog we also then present the lower and upper boundaries of the 80\% HPD CI peaks (i.e. where the $P(z)$ crosses the threshold).
We refer the interested reader to Fig.~1 of \citet{2016MNRAS.457.4005W}, for a more detailed explanation and illustration of the concept as well as an excellent discussion on the motivation behind such a treatment of redshift posteriors.

For our measure of robust scatter, we use the normalised median absolute deviation,  $\sigma_{\textup{NMAD}}$, defined as:
\begin{equation}
\sigma_{\textup{NMAD}} =1.48 \times \text{median} ( \left | \Delta z \right | / (1+z_{\textup{spec}})),
\end{equation}
where $\Delta z = z_{1, \textup{median}} - z_{\textup{spec}}$.
Similarly, we define outliers as 
\begin{equation}
	\left | \Delta z \right | / (1+z_{\text{spec}}) > 0.15,
\end{equation}
as is common for the literature \citep[e.g. ][]{Dahlen:2013eu}.
In Table~\ref{tab:stats_all} we present the $\sigma_{\textup{NMAD}}$ and the outlier fraction ($\textup{OLF}$) for the full spectroscopic redshift sample and the various subsets defined by our broad multi-wavelength selection criteria from Section~\ref{sec:mw_class}.
Statistically we see confirmation of the qualitative picture discussed above.
Photo-$z$s for the non-AGN selected population are excellent, with very low scatter and a low outlier fraction.
Both scatter and outlier fraction for the AGN selected subsets are significantly worse, with the overall outlier fraction being $\sim 30\%$ for this sample.

\begin{table}
\caption{Photo-$z$ quality metrics for the full redshift sample, the LOFAR-detected spectroscopic redshift sample and the various subsets of both samples defined by our multi-wavelength classification (Section~\ref{sec:mw_class}).}
\centering
\begin{tabular}{lccc}
Subset & $N$ & $\sigma_{\textup{NMAD}}$ & $\textup{OLF}$ \\
\hline
\hline
\multicolumn{4}{c}{Full spectroscopic sample} \\
\hline
All & 314625 & 0.041 & 0.104 \\
Galaxies & 233002 & 0.031 & 0.034 \\
AGN & 81623 & 0.123 & 0.306 \\
\hline
QSOs & 69251 & 0.110 & 0.274 \\
Spectroscopic AGN & 75854 & 0.123 & 0.306 \\
X-ray AGN & 1689 & 0.070 & 0.132 \\
IR AGN & 34527 & 0.083 & 0.169 \\
\hline
& & & \\
& & & \\
\hline
\multicolumn{4}{c}{LoTSS spectroscopic sample} \\
\hline
All & 29535 & 0.039 & 0.079 \\
Galaxies & 21133 & 0.031 & 0.015 \\
AGN & 8402 & 0.090 & 0.241 \\
\hline
QSOs & 7025 & 0.084 & 0.221 \\
Spectroscopic AGN & 6811 & 0.102 & 0.266 \\
X-ray AGN & 669 & 0.060 & 0.135 \\
IR AGN & 5336 & 0.090 & 0.220 \\
\end{tabular}
\label{tab:stats_all}
\end{table}

When restricting the analysis to sources that are detected in the LoTSS radio catalog (Table~\ref{tab:stats_all}), the picture is very similar but performance is generally better.
Scatter for the non-AGN selected population is unchanged and there is significant improvement in $\textup{OLF}$ with a reduction to 1.3\%.
Across the AGN subsets there is a significant improvement in both metrics, although we note the infrared selected AGN performance is slightly worse for the radio detected sample.
The improved performance for radio-detected sources (at least at lower redshifts, e.g. $z \lesssim 1$) mirrors that observed in \citetalias{Duncan:2017wu} and can partly be attributed to the fact that radio sources are typically hosted by the most massive (and hence brightest) galaxies \citep{2014ARA&A..52..589H} and will typically have higher signal-to-noise than the general galaxy population at the same redshift.

\subsection{Photometric redshift statistics as a function of redshift and magnitude}
\begin{figure*}
\centering
  \includegraphics[width=1\columnwidth]{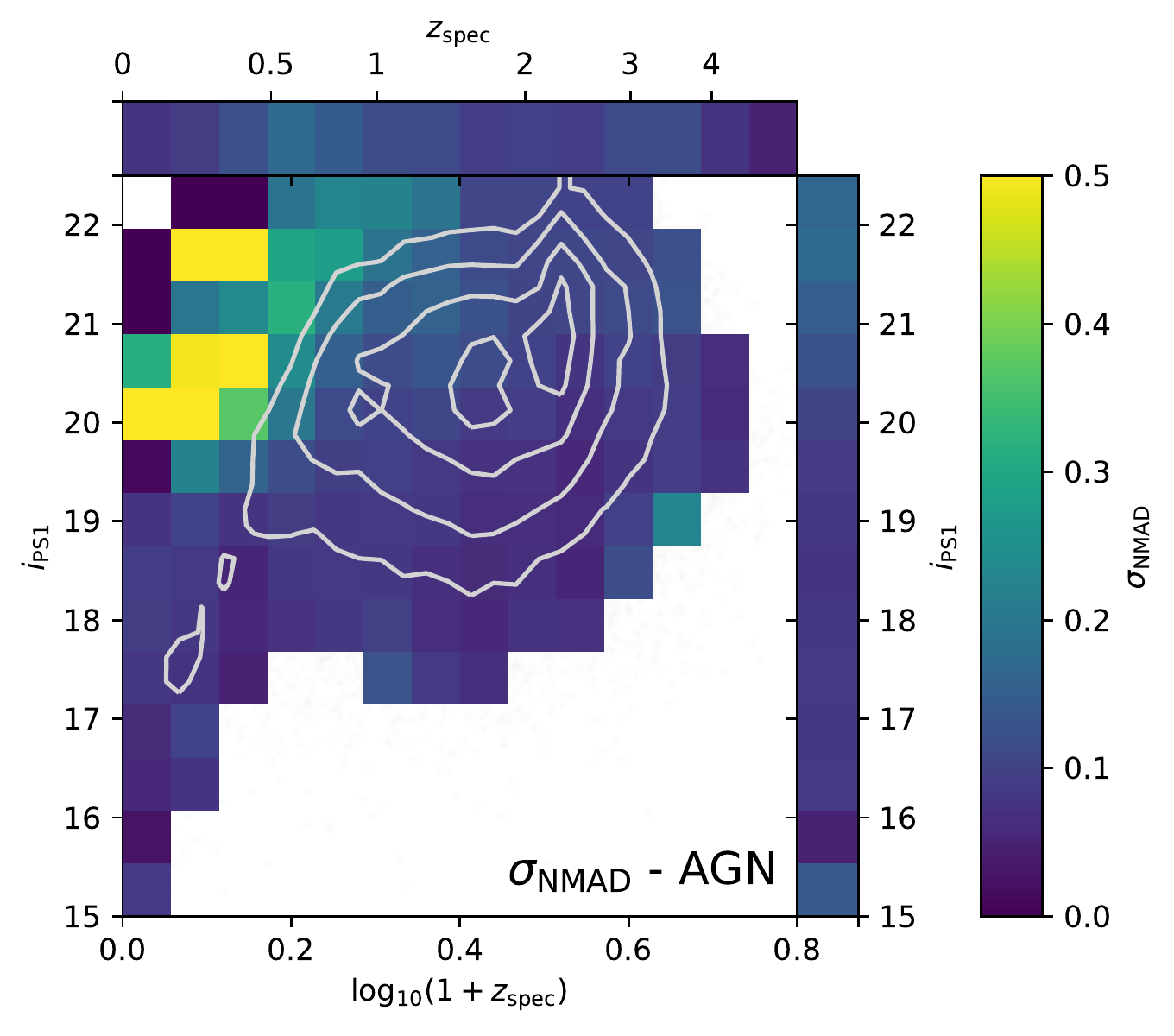}
    \includegraphics[width=1\columnwidth]{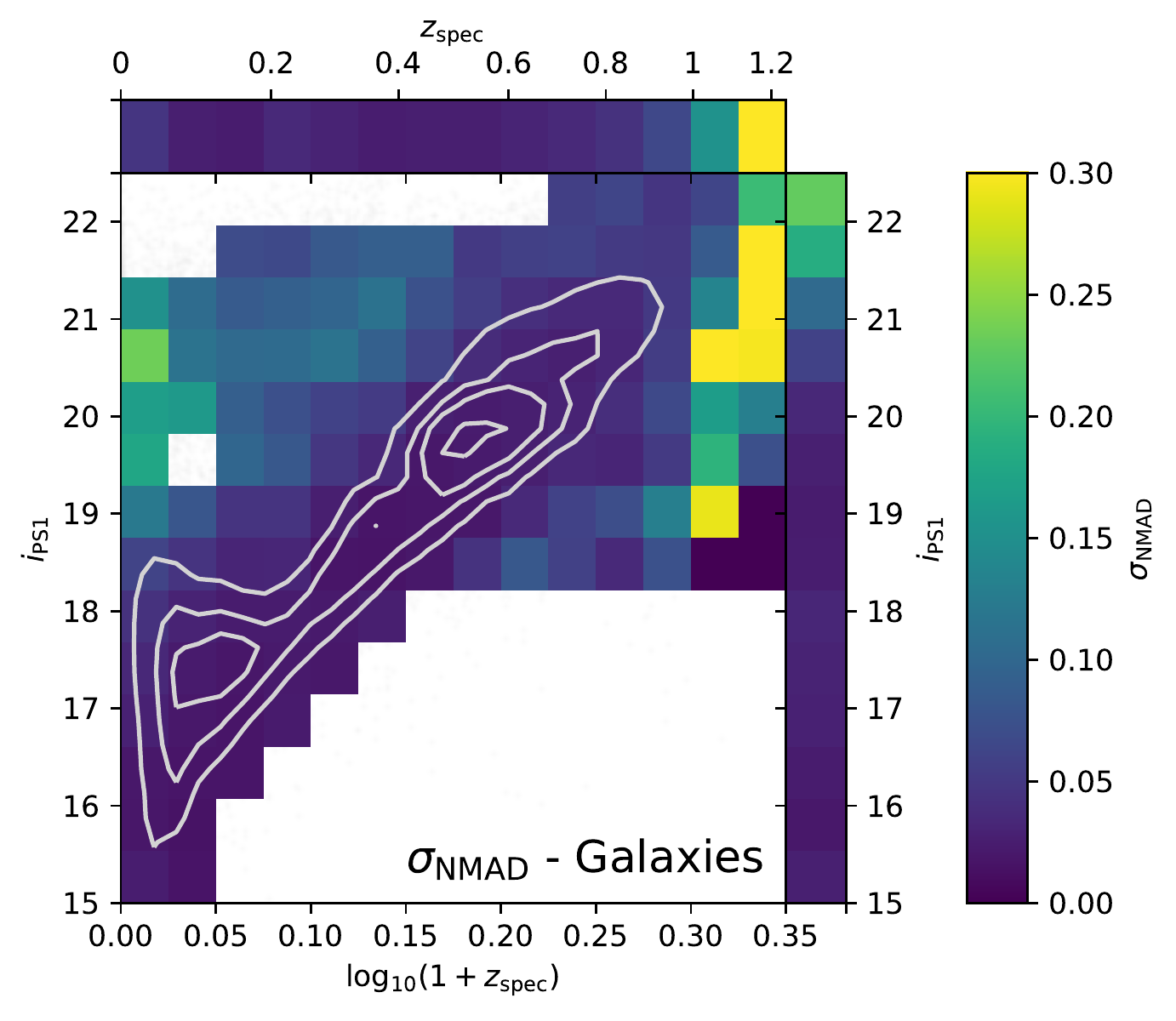}
  \includegraphics[width=1\columnwidth]{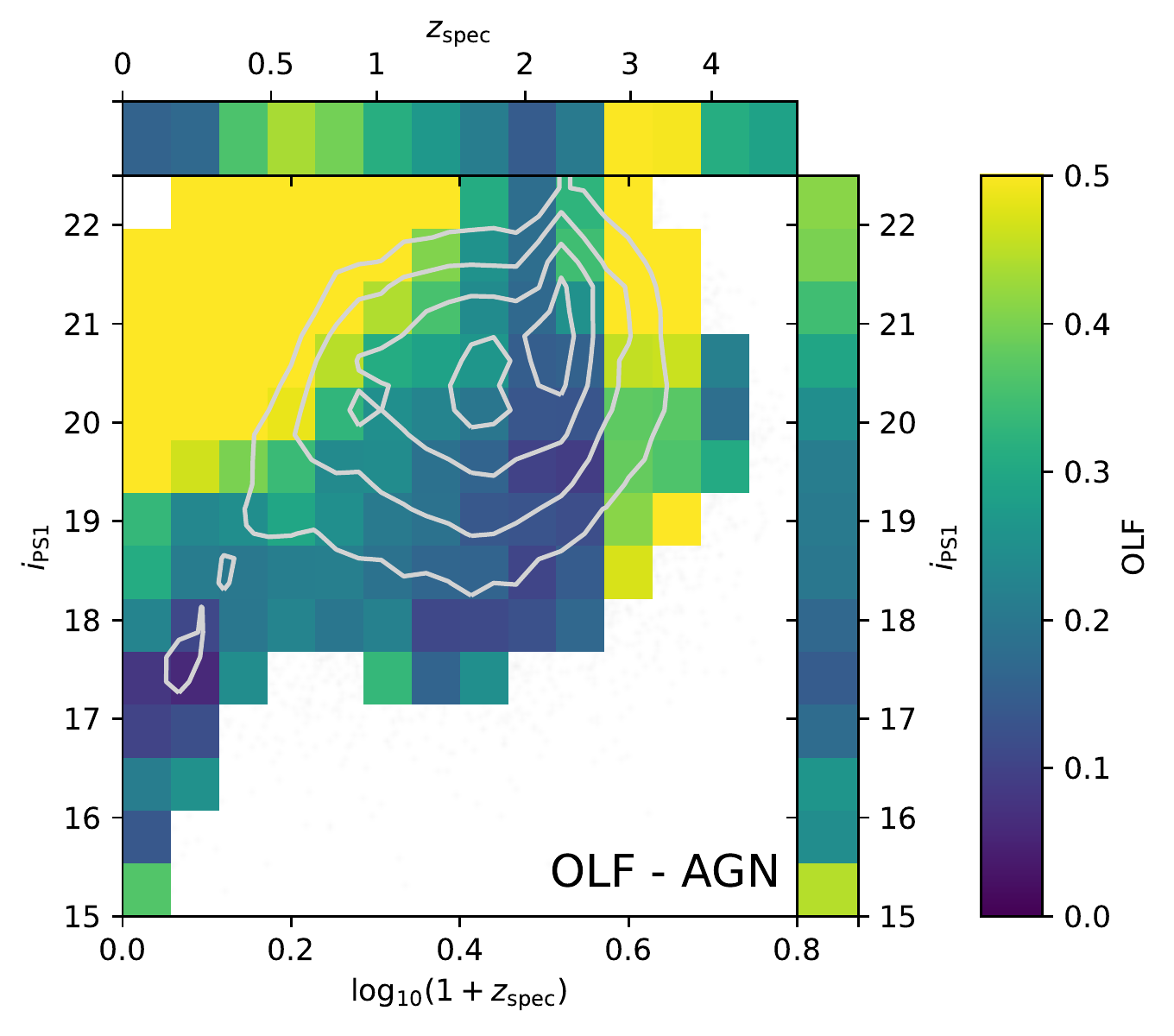}
  \includegraphics[width=1\columnwidth]{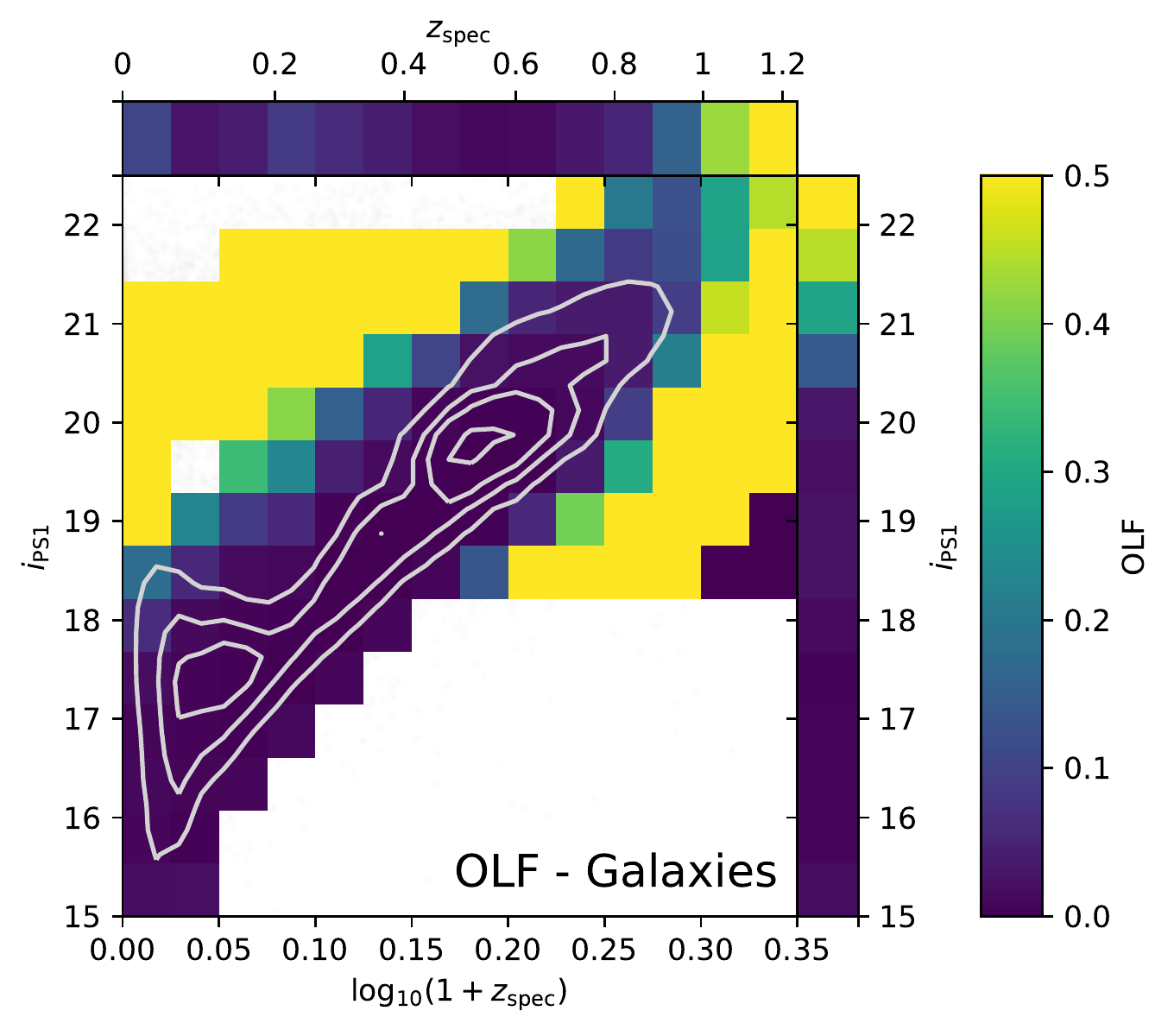}

  \caption{Robust scatter ($\sigma_{\textup{NMAD}}$; upper panels) and outlier fraction ($\textup{OLF}$; lower panels) for the consensus photo-$z$ estimate ($z_{1,\textup{median}}$) as a function of spectroscopic redshift and apparent $i_{\textup{PS1}}$ magnitude. For the AGN subset (left panels) each cell corresponds to a minimum of 30 sources with the colour of the cell representing the scatter of that subset. For the galaxy plots (right panels) each cell corresponds to a minimum of 100 sources. The top and side bar of each panel shows the trends in $\sigma_{\textup{NMAD}}$ or  $\textup{OLF}$ averaged over all magnitudes and redshifts respectively. For reference, we also plot the distribution of the spectroscopic training sample within this parameter space as grey contours -  plotted contours are linearly spaced in source density.}
  \label{fig:z_mag_scatter_olf}
\end{figure*}

In the upper panels of Fig.~\ref{fig:z_mag_scatter_olf} we show the measured robust scatter for the AGN and galaxy subsets in bins of both $z_{\textup{spec}}$ and apparent optical magnitude ($i_{\textup{PS1}}$).
The lower panels of Fig.~\ref{fig:z_mag_scatter_olf} presents the corresponding OLF over the same parameter space.
Additionally, for all four of the diagnostic plots we also present the relative density of the spectroscopic sample within this parameter space for reference.

From these figures we can see that the photo-$z$ estimates are in general very good within the regime for which a large number of spectroscopic sources exist.
Typical scatter for the AGN population is $\sigma_{\textup{NMAD}} \approx 0.1$, comparable to or better than other estimates for similar populations in the literature \citep[e.g. ][]{Richards:2001ct,Brodwin:2006dp,2012MNRAS.424.2876M,Chung:2014it}.
 Furthermore these estimates are also better at $1 < z < 3$ than photo-$z$ estimates calculated using the same method on deeper photometric samples \citepalias{Duncan:2017wu,Duncan:2017ul}.
 We attribute this performance to the larger training sample for these source types used in this work, leading to excellent \textsc{GPz} performance in this regime.

Outlier fractions for the AGN population follow a similar trend but with generally slightly poorer performance.
In the regions of parameter space that are sparsely sampled by the spectroscopic sample, the outlier fraction rapidly deteriorates.
However, between $0 < z < 2.5$ the outlier fraction averaged over the whole AGN population remains good enough for many science cases - especially when taking into account the larger samples that are now available compared to studies that only make use of the spectroscopic sample.

For the galaxy population, we find the outlier fraction for the bulk of the parameter space between $0 < z< 0.8$ to be exceptional - with outlier fractions at the sub-percent level for some redshifts and magnitudes.
We can also now see more quantitatively the previously observed fall-off in photo-$z$ accuracy (Fig.~\ref{fig:z_mag_scatter_olf} upper panels) and reliability (Fig.~\ref{fig:z_mag_scatter_olf} lower panels) at $z > 0.8$.

\subsection{Photo-$z$ properties for the LOFAR detected population}
Finally, we explore the quality of the consensus photo-$z$ estimates as a function of their radio properties.
In \citetalias{Duncan:2017wu}, we found that for the template-only estimates there was a weak trend such that more luminous radio sources typically had poorer photo-$z$ performance.
However, the available spectroscopic sample for radio-detected sources was too small to provide robust conclusions on what was responsible for this trend.
In \citetalias{Duncan:2017ul}, we illustrated how the inclusion of \textup{GPz} estimates for the AGN population results in significant improvements for the most luminous radio sources.

\begin{figure}
\centering
  \includegraphics[width=\columnwidth]{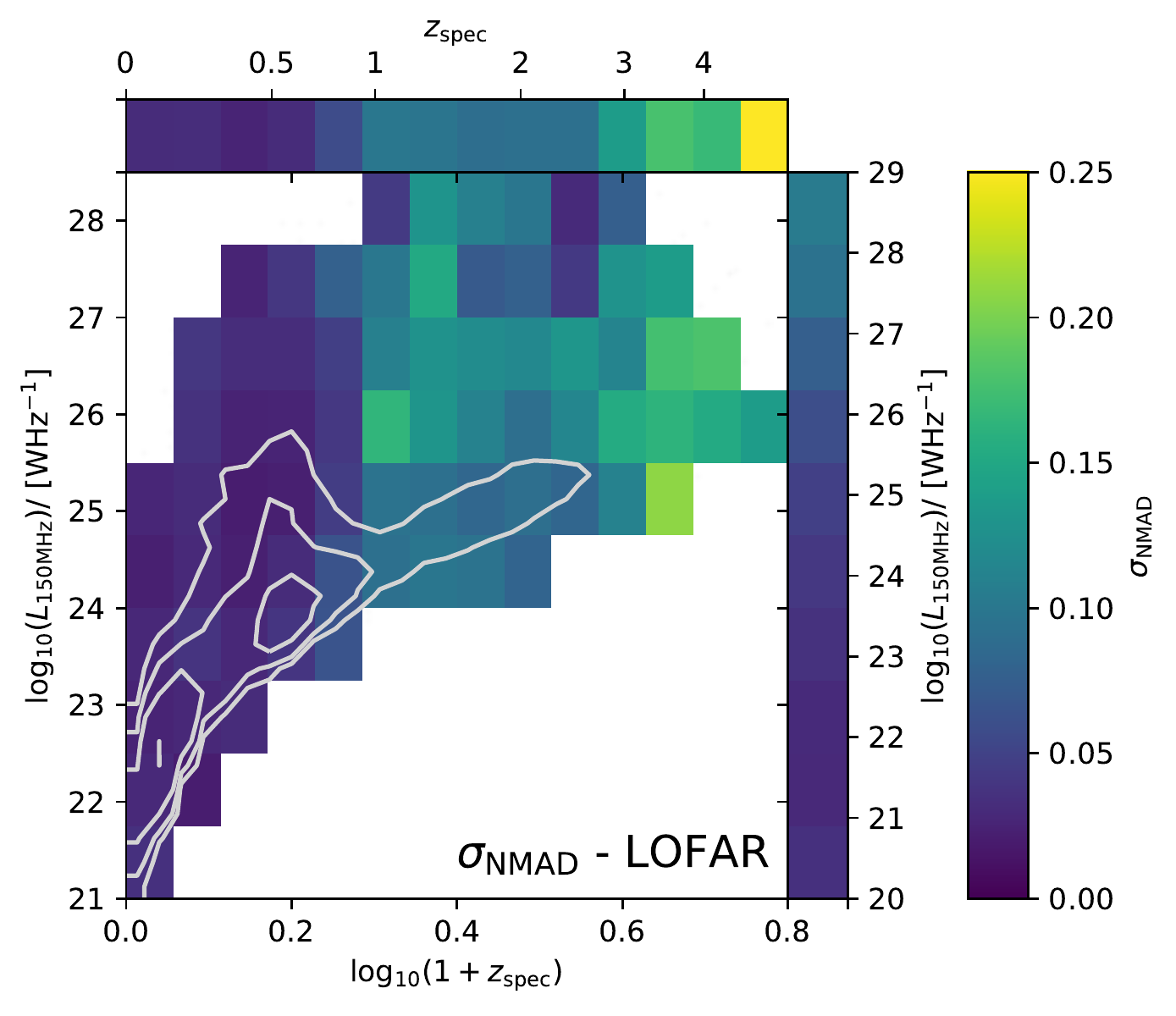}
  \includegraphics[width=\columnwidth]{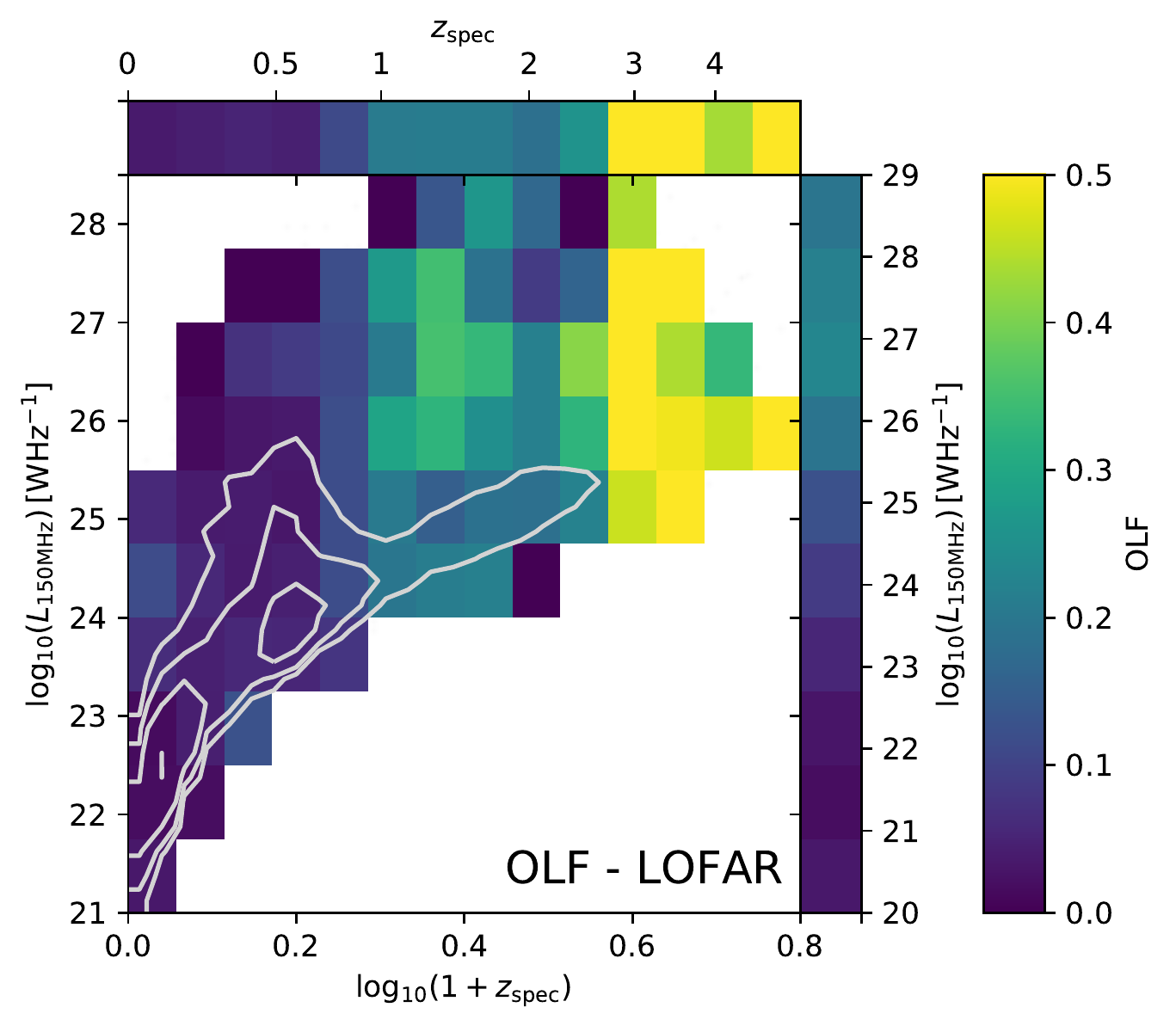}
  \caption{Robust scatter ($\sigma_{\textup{NMAD}}$; top) and outlier fraction ($\textup{OLF}$; bottom) for the consensus photo-$z$ estimate as a function of spectroscopic redshift and 150 MHz radio continuum luminosity. Each cell corresponds to a minimum of 100 galaxies. The top and side panel show $\sigma_{\textup{NMAD}}$ show the trends averaged over all magnitudes and redshifts respectively. For reference, we also plot the distribution of the spectroscopic training sample within this parameter space.}
  \label{fig:specz_l150_nmad_olf}
\end{figure}

The unprecedented sample of radio sources presented in \citetalias{Shimwell:2018to} and their reliable optical counterparts \citepalias{Williams:2018us} means that we are now able to explore these trends with much greater precision. 
In \citetalias{Williams:2018us}, Pan-STARRS or WISE cross-identifications were found for 71\% of the LoTSS radio sources. 
Following the method described in this work and the additional selection criteria required for photo-$z$ estimation (i.e. PS1 $g$, $r$ and $i$ detections at a minimum), we are able to estimate photo-$z$ for 70\% of the sources with optical IDs.
We are therefore able to provide photo-$z$ estimates for $49.5\%$ of the LOFAR sources presented in DR1.
When including additional spectroscopic redshifts that did not satisfy the stricter requirements for training \textup{GPz} alongside the spectroscopic training sample, a total of 29535 of the LoTSS sample have spectroscopic redshifts.

In Fig.~\ref{fig:specz_l150_nmad_olf} we present the $\sigma_{\textup{NMAD}}$ and $\textup{OLF}$ as a function of spectroscopic redshift and 150MHz radio luminosity.
As in the figure in the previous section, we also over-plot the distribution of spectroscopic sources within this parameter space for reference.
When converting from observed flux density to rest-frame radio luminosity, we assume an average spectral slope of $\alpha = -0.7$ for all sources.

Within a given spectroscopic redshift bin, we see no evidence for any significant trend with radio luminosity in either the scatter or outlier fraction.
Instead, it is clear that the previously observed trends can be attributed solely to the trends as a function of redshift.
For both metrics we see a clear evolution with $z_{\textup{spec}}$, such that the scatter and outlier fractions for the highest redshift sources are significantly worse than for sources with similar radio luminosity at low redshift.
This trend may be driven by either selection effects within the spectroscopic sample or evolution in the radio population itself (or likely some combination of the two).
However, we leave that question for subsequent studies to investigate.

\section{Rest-frame Properties}\label{sec:derivedquantities}
Ultimately, for all sources within the LoTSS HETDEX field we would like to know the physical properties of the host galaxies, including constraints on the relative contributions to the optical spectral energy distribution (SED) from stellar or accretion emission processes.
While full panchromatic SED fitting codes such as \textsc{AGNFitter} \citep{2016ApJ...833...98C} mean that it is possible to disentangle these different components and characterise radio sources \citep[e.g.][]{2017MNRAS.469.3468C,2018MNRAS.475.3429W}, the scale of the LoTSS DR1 sample and the more limited multi-wavelength data available mean that such measurements are beyond the scope of this data release.
However, while such detailed fits and the corresponding physical properties are desirable, much can be learned from the rest-frame colours and magnitudes of sources. 
For the full sample of LOFAR selected sources with optical counterparts and photometric (or spectroscopic) redshifts, we therefore estimate a broad range of rest-frame magnitudes.

We estimate rest-frame magnitudes using the template interpolation feature of the \textsc{Eazy} photometric redshift code \citep{Brammer:2008gn}.
Fixing the redshift to the best available redshift estimate ($z_{\textup{spec}}$ where available, $z_{1, \textup{median}}$ otherwise) we re-fit all radio sources using all three template libraries.
Rest-frame magnitudes can then be calculated based on the flux in a given filter for the best-fitting template observed at $z=0$.
When re-fitting the SEDs for rest-frame magnitudes, we make use of the forced Kron fluxes for the PanSTARRs  $g,r,i,z,y$ optical bands and the profile-fitting magnitudes for WISE W1-3 bands.

In addition to the observed bands used in the photo-$z$ fitting, rest-frame magnitudes were also estimated in additional filters common in the literature.
Specifically, we calculate magnitudes for SDSS $u,g,r,i,z$, Johnson-Cousins $U, B, V$ and $I$, and the 2MASS $J$ and $K_{s}$ near-infrared filters. 

For any source that satisfies one or more of the AGN selection criteria, we use the rest-frame optical and near-infrared magnitudes from the XMM-COSMOS \citep{Salvato:2008ef} fits and the mid-infrared rest-frame magnitudes from the `Atlas of Galaxy SEDs' \citep{Brown:2014jd} fits.
For the remaining sources, values are taken from the fits to the \textsc{Eazy} templates - with the exception of the WISE filters that we take from the `Atlas' estimates.

\begin{figure*}
\centering
  \includegraphics[width=0.8\paperwidth]{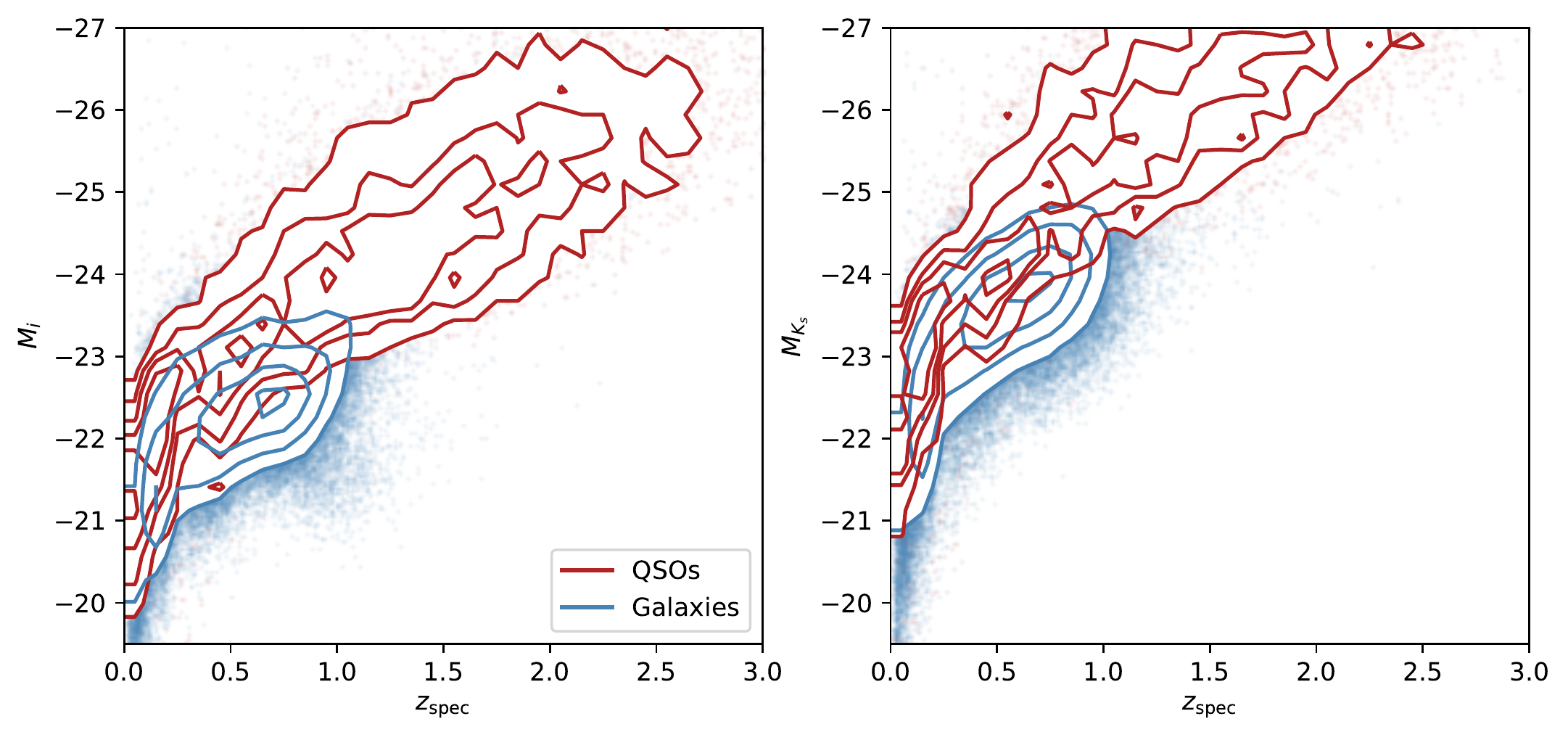}
  \caption{Estimated rest-frame $i$ (left panel) and $K_{S}$ (right panel) magnitudes as a function of redshift for two subsets of the LOFAR detected population - optically selected QSOs (red distribution) and sources that are not selected as optical, X-ray or infrared AGN.}
  \label{fig:rf_vs_redshift}
\end{figure*}

In Fig.~\ref{fig:rf_vs_redshift} we plot the estimated $i$ and $K_{S}$ magnitudes as a function of redshift (spectroscopic where available, photometric otherwise) for two subsets of the radio population to illustrate the typical range and distribution of rest-frame magnitudes within the sample.

\begin{figure*}
\centering
  \includegraphics[width=0.8\paperwidth]{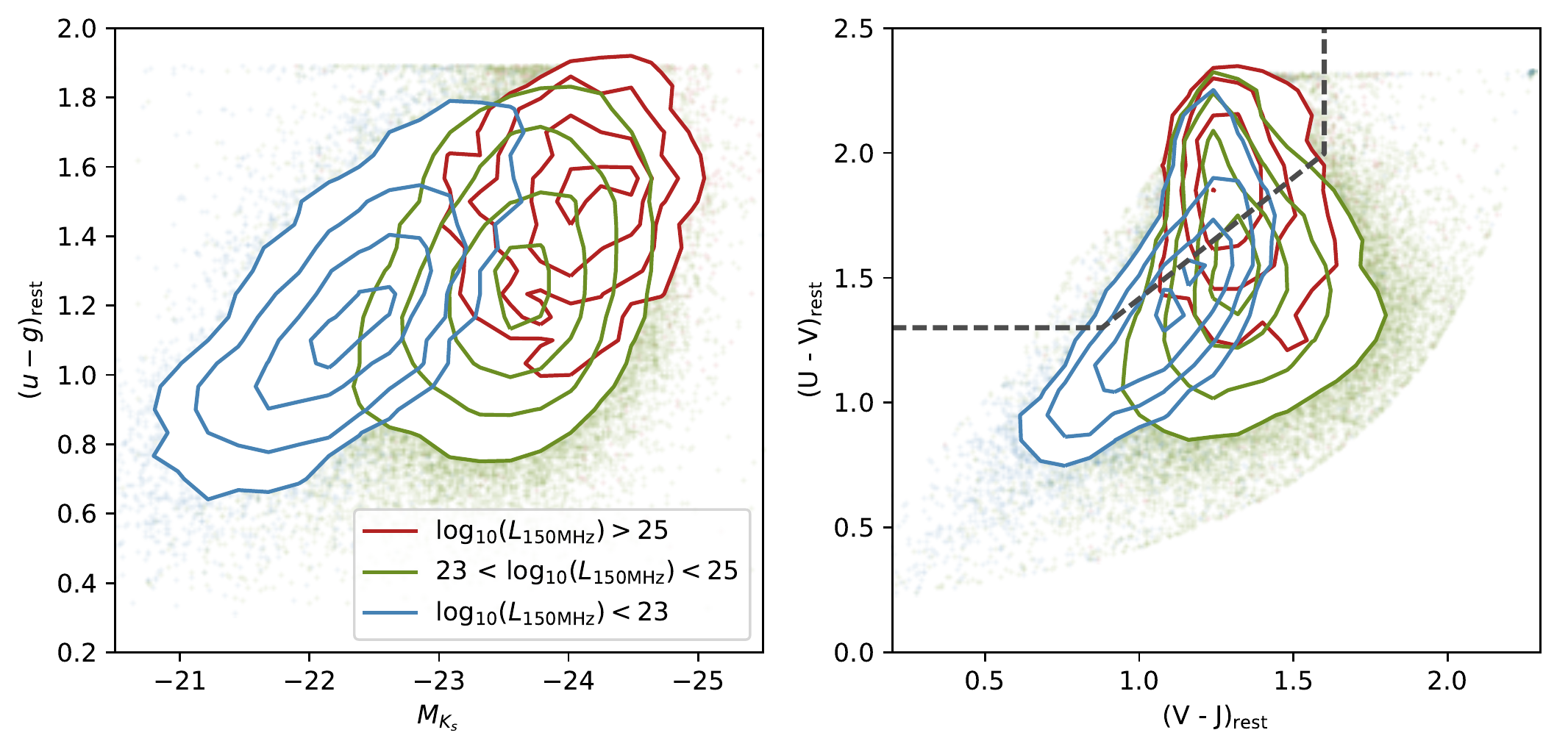}
  \caption{Left: Observed colour-magnitude distribution, $u-g$ vs $M_{K_{s}}$ of the $0.1 < z < 0.8$ in three bins of radio luminosity for the LOFAR selected sources that are not classified as optical, IR or X-ray AGN. The most luminous radio sources are hosted in galaxies that are red and bright in the near-IR (a strong proxy for stellar mass). Right: Rest-frame $U - V$ vs $V - J$ optical colours for the LOFAR detected population for the same bins in radio luminosity. The black dashed line shows the typical boundary used to separate star-forming and quiescent stellar populations \citep[e.g. ][]{2009ApJ...691.1879W}.
}
  \label{fig:rf_colour}
\end{figure*}

Additionally, to provide further validation of the rest-frame magnitudes and illustrate their scientific potential, in Fig~\ref{fig:rf_colour} we show two optical diagnostic plots used commonly in the literature.
The left-panel of Fig~\ref{fig:rf_colour} shows the observed colour-magnitude distribution of the $0.1 < z < 0.8$ LOFAR selected population in bins of radio luminosity.
In addition to the selection in redshift range, a simple redshift quality cut based on the posterior uncertainties is applied such that in Fig~\ref{fig:rf_colour} we plot only sources where $(0.5 \times \left | z_{1,\textup{max}} - z_{1,\textup{min}}\right | / (1 + z_{1,\textup{median}})) < 0.2$, or a spectroscopic redshift is available (yielding a sample of 78735 radio sources).
We can clearly see that the most luminous radio sources tend to reside in galaxies that are more luminous in the near-IR (with $M_{K_{s}}$ a good proxy for stellar mass) and have very red optical colours; consistent with expectations for radio-loud AGN \citep{2014ARA&A..52..589H}.
In contrast, the lower luminosity radio population is hosted in galaxies that are typically bluer and lower mass.

In the right-panel of Fig~\ref{fig:rf_colour} we show the distribution of the same sources in the widely used $UVJ$ colour diagram \citep{2009ApJ...691.1879W}.
This plot illustrates that not only are the most luminous radio sources found in red galaxies, but those galaxies are likely red due to having old quiescent stellar populations.
Conversely, lower luminosity radio sources have optical colours more consistent with star-forming galaxies.

As mentioned above, these trends are well established for the low-redshift radio populations.
However, the unprecedented depth, sensitivity and size of the LoTSS DR1 catalog mean that such trends can now be explored in much greater detail, extending to higher redshifts or lower radio luminosities than previously possible for such a large statistical sample. 
The redshifts and rest-frame properties presented in this paper are intended to enable such studies and many others besides.

\section{Final catalog}\label{sec:catalog}
The catalog presented in this work builds upon both the radio \citetalias{Shimwell:2018to} and optical identification LoTSS DR1 catalogs described in \citetalias{Williams:2018us}.
The contents of the catalog added by this work are as follows:
\begin{itemize}
	\item The best available redshift for a given source, `$z_{\textup{best}}$', where spectroscopic redshift is used if available and the best available photo-$z$ ($z_{1, \textup{median}}$) used otherwise.\footnote{Due to the conservative selection used to define the spectroscopic training sample, the included $z_{\textup{spec}}$ included in the sample are not explicitly intended to be complete. Additional spectroscopic redshifts may therefore be available within the wider literature.}.
	\item The source of the best available redshift, `$z\_{\textup{best}}\_{\textup{source}}$' where 1 corresponds to spectroscopic redshift and 0 corresponds to the photo-$z$ presented in this work.
	\item Median of the primary redshift peak, `$z_{1, \textup{median}}$'. This is the `best' estimate of the photo-$z$ from this work.
	\item Lower (`$z_{1, \textup{min}}$') and upper (`$z_{1, \textup{max}}$') bounds of the primary 80\% HPD CI peak (where the redshift $P(z)$ crossed the credible interval).
	\item Fraction of the integrated probability included in the primary peak contained within the 80\% HPD CI, `$z_{1, \textup{area}}$' ($\leq 0.8$ by definition).
	\item Properties of the secondary 80\% HPD CI peak if it exists:  `$z_{2, \textup{median}}$', `$z_{2, \textup{min}}$, `$z_{2, \textup{max}}$' and `$z_{2, \textup{area}}$'.
\end{itemize}

Also included for all sources are the multi-wavelength AGN classifications used during photo-$z$ estimation.
\begin{itemize}
\item `specAGN':	Flag indicating spectroscopically identified AGN (1 = AGN).
\item `mqcAGN': Flag indicating whether source is included in Million Quasar Catalog compilation \citep{2015PASA...32...10F}, where 1 means a source is included.
\item `XrayClass': 2RXS or XMMSL2 X-ray source class - 0 = WISE source, but no X-ray match, 1 = AGN, 2 = Galaxy/Star \citep[based on criteria in]{Salvato:2017gj}.
\item `2RXS\_ID': ID in 2RXS catalog (if available)
\item `XMMSL2\_ID':	ID in XMMSL2 catalog (if available)
\item `IRClass': Bit-flag indicating WISE AGN Class based on \citet{2013ApJ...772...26A} selection criteria, where 1 = 90\% completeness criteria, 2 = 75\% completeness criteria, 4 = 75\% reliability criteria and 8 = 90\% reliability criteria.
\end{itemize}

Finally, for all sources for which a redshift estimate exists (either spectroscopic or photometric), we include the additional rest-frame magnitudes presented in Section~\ref{sec:derivedquantities}:
\begin{itemize}
	\item Estimated rest-frame magnitudes in the SDSS filter set,  `$X\_\textup{sdss}\_\textup{rest}$', where $X = u,g,r,i$ or $z$. 
	\item Estimated rest-frame magnitudes in the reference Johnson-Cousins optical filters, `$X\_\textup{rest}$' where $X = U, B, V$ or $I$.
	\item Estimated rest-frame magnitudes in the 2MASS $J$ (`$J\_\textup{rest}$') and $K_\textup{s}$ (`$K\_\textup{rest}$') near-infrared filters. 
	\item Estimated WISE rest-frame magnitudes - `$W1\_\textup{rest}$', `$W2\_\textup{rest}$' and `$W3\_\textup{rest}$'.
\end{itemize}

\section{Future prospects}
The photo-$z$ estimates presented in this work make use of the best all-sky photometric datasets and the latest techniques to provide the best estimates practical for the large area.
However, future data releases of the LoTSS survey will be able to exploit both improved photometric datasets and greatly enhanced photo-$z$ techniques and training samples; resulting in greater fraction of optical cross-identifications for LoTSS sources and more accurate photo-$z$ and physical parameter estimates.

Under the umbrella of the NOAO Legacy Surveys program, new photometry reaching depths $\sim 1$ magnitude deeper than the PanSTARRS $3\pi$ survey in the $g$, $r$ and $z$ bands will soon be available 
At declinations of $\gtrsim 30\deg$ these observations are provided by the combination of the Beijing-Arizona Sky Survey \citep[BASS;][]{2017AJ....153..276Z} and the Mayall $z$-band Legacy Survey \citep[MzLS;][]{2016AAS...22831702S}.
At lower declination, the corresponding $g$, $r$ and $z$ is provided by the Dark Energy Camera Legacy Survey (DeCALS; PI: D. Schlegel and A. Dey).

A key advantage of the catalogs provided by these surveys is the inclusion of optical prior driven deconfusion of the unWISE data release of WISE photometry.
The unWISE processing maintains the native resolution of the shorter wavelength WISE bands and incorporating the additional W1 and W2 observations provided by the post-cryogenic WISE mission (NEOWISE ). 
For the input optical prior sources, the model fitting photometry is able to provide robust measurements to significantly deeper magnitudes than reached by the AllWISE catalogs used in this work.
Although there are fewer optical bands available from BASS+MzLS or DeCALS (compared to PS1) for photo-$z$ estimation or physical modelling, the improvement to the \textup{GPz} estimates shown when W1 is included in the fitting (e.g. see Fig.~\ref{fig:gpz_gal_speczphotz}) suggests that the availability of WISE constraints for \emph{all} optical sources will result in improved photo-$z$ estimates.

Furthermore, ongoing photometric surveys at complementary wavelengths will likely greatly enhance the available datasets over the LoTSS regions.
For example, the Canada France Hawaii Telescope `Legacy for the $u$-band all-sky universe' survey (CFHT-Luau; PI: McConnachie) aims to reach a depth of $\sim 24.2$ over $> 4000$ sq.deg in the northern hemisphere.
Additionally, the UKIRT Hemisphere Survey \citep[UHS;][]{2018MNRAS.473.5113D} can fill the significant gap in wavelength coverage in the near-infrared by providing $J$ band observations over a similar area in the northern sky.

Commencing in 2019, the WEAVE-LOFAR spectroscopic survey \citep{Smith:2016vw} will obtain $\gtrsim 10^{6}$ spectra of LOFAR selected radio sources over the northern hemisphere.
In addition to providing robust spectroscopic redshifts for a significant number of the most luminous radio sources for which photo-$z$s are particularly difficult (see Fig.~\ref{fig:specz_l150_nmad_olf}), WEAVE-LOFAR will provide an unprecedented spectroscopic training sample for the LoTSS radio population.

Specifically, the deepest tier of WEAVE-LOFAR aims to provide spectroscopic observations for radio sources down to flux densities of $S_{\nu, 150\textrm{MHz}} = 100\mu$Jy over $\sim50$sq.deg in deep fields -- the detection limit of the LoTSS survey.
The nature of the radio selection (as compared to optical selection of most available spectroscopic surveys) means the the WEAVE-LOFAR data will provide extensive training samples in regions of magnitude/colour/redshift space that are currently not well sampled.
As illustrated in Fig.~\ref{fig:z_mag_scatter_olf}, it is in these regions of parameter space, where few training sources are present, that the \textup{GPz} estimates are poorest. 

Furthermore, within the region of sky presented in this work, the HETDEX survey itself will provide large numbers of additional spectroscopic sources for blind detections of both redshift Lyman-alpha emitters (radio-loud/-quiet quasars etc.) and $O_{\textsc{iii}}$/$O_{\textsc{ii}}$ emitters at low-redshift.
It is not currently known what fraction of emission line selected galaxies will be detected by the LoTSS survey.
But if even only a small fraction ($<1-5$) of the line-emitters are also LOFAR sources, the number of spectroscopic redshifts provided by the HETDEX survey will be substantial.
By leveraging these improved training samples and the deeper photometry available, the photo-$z$ precision and reliability for future LoTSS data releases will greatly improved.

Finally, in light of the photometric redshift quality achieved for a new generation of radio continuum survey (LoTSS), it is worth exploring what these results may mean for future radio continuum surveys such as the Square Kilometre Array (SKA).
Given the anticipated depths of the SKA continuum surveys and the rapid rise in continuum source counts, the majority of SKA sources will be fainter radio sources than those presented in this data release \citep{2015aska.confE..67P,2015aska.confE..18J}. 
In \citet{Duncan:2017wu}, we illustrated that the typical photo-$z$ accuracy (for template estimates) improves as one probes fainter radio sources. 
Improvements to the photo-$z$ estimates for bright source populations yielded by the suggestions presented above will yield comparable results for future SKA observations.

As a southern hemisphere survey, the SKA stands to benefit from substantially deeper all-sky optical photometry than will be available in the north thanks to the Large Synoptic Survey Telescope \citep[LSST;][]{2012arXiv1211.0310L}.
The combination of deep LSST optical photometry and complementary near-infrared observations from the Euclid survey mission \citep{2011arXiv1110.3193L} will likely yield dramatically improved results when combined with methods such as that presented in this study.
As illustrated by previous studies such as \cite{2012MNRAS.424.2876M}, the inclusion of multiple near-infrared bands in photo-$z$ estimates can yield significant improvements for difficult source types such as quasars.

Although the SKA will not benefit from the full extent of spectroscopy provided by WEAVE-LOFAR, photo-$z$ estimates for SKA sources will be able to exploit the WEAVE-LOFAR DEEP and WIDE samples in equatorial regions to provide representative training samples for machine learning estimates \citep[see][for further details]{Smith:2016vw}.
Additional spectroscopy for large numbers of SKA detected sources will likely also be provided by optical and X-ray selected spectroscopic campaigns with instruments such as 4MOST \citep{2016SPIE.9908E..90S,2017mbhe.confE..76B}.
Given these available photometric and spectroscopic training samples, the prospects for attaining photo-$z$ precision and reliability suitable for tackling the science goals of the SKA look promising.

\section{Summary}\label{sec:summary}
In this paper we present details of photometric redshift (photo-$z$) estimates estimates produced for the LOFAR Two-meter Sky Survey (LoTSS) First Data Release.
Photo-$z$ are estimated for the full $\sim 400$ deg$^{2}$ of the HETDEX Spring Field using optical photometry from the PanSTARRS $3\pi$ Survey and mid-infrared photometry from the Wide-field Infrared Survey Explorer mission \citep[WISE;][]{Wright:2010in}.
Our photo-$z$ method combines multiple traditional template fitting and empirical training based estimates to produce an optimised consensus redshift prediction and includes explicit efforts to calibrate the photo-$z$ uncertainties.

Averaged over all 314625 sources in the spectroscopic training and test sample, the resulting consensus photo-$z$s have a robust scatter of $\sigma_{\textup{NMAD}} =  0.041$ and an outlier fraction of 0.104.
However, there is strong variation in photo-$z$ quality as a function of optical source type, with sources that satisfy one or more AGN selection criteria (optical, X-ray or infrared) having a significantly worse performance ($\sigma_{\textup{NMAD}} = 0.123$ and $\textup{OLF} = 0.306$) than those that do not (0.031 and 0.034 respectively).
Additionally, as a function of both optical magnitude and spectroscopic redshift there are strong trends for both subsets of the population.

For the LoTSS DR1 radio detected sample, we find that at a given redshift, there is no strong trend in photo-$z$ quality as a function of radio luminosity.
However there is clear deterioration in photo-$z$ quality as function of redshift for a given radio luminosity that we attribute to selection effects in the spectroscopic sample and/or intrinsic evolution within the radio population.
Finally we present details of rest-frame optical and mid-infrared magnitudes for the LoTSS DR1 sample estimated using our consensus photometric redshift estimates (or spectroscopic redshift when available).

\section*{Acknowledgements}
The authors thank the anonymous referee for their time and efforts as well as their prompt response in helping to improve the manuscript.
This paper is based on data obtained with the International LOFAR Telescope (ILT) under project codes LC2\_038 and LC3\_008. LOFAR \citep{vanHaarlem:2013gi} is the Low Frequency Array designed and constructed by ASTRON. It has observing, data processing, and data storage facilities in several countries, which are owned by various parties (each with their own funding sources), and which are collectively operated by the ILT foundation under a joint scientific policy. The ILT resources have benefited from the following recent major funding sources: CNRS-INSU, Observatoire de Paris and Universit\'{e} d’Orl\'{e}ans, France; BMBF, MIWF-NRW, MPG, Germany; Department of Business, Enterprise and Innovation (DBEI), Ireland; NWO, The Netherlands; The Science and Technology Facilities Council (STFC), UK.

The data used in work was in part processed on the Dutch national e-infrastructure with the support of SURF Cooperative through grant e-infra 160022 \& 160152. The LOFAR software and dedicated reduction packages on \url{https://github.com/apmechev/GRID\_LRT} were deployed on the e-infrastructure by the LOFAR e-infragroup, consisting of J. B. R. Oonk (ASTRON \& Leiden Observatory), A. P. Mechev (Leiden Observatory) and T. Shimwell (ASTRON) with support from N. Danezi (SURFsara) and C. Schrijvers (SURFsara).  

This research has made use of data analysed using the University of Hertfordshire high-performance computing facility (\url{http://uhhpc.herts.ac.uk/}) and the LOFAR-UK computing facility located at the University of Hertfordshire and supported by STFC [ST/P000096/1].

The Pan-STARRS1 Surveys (PS1) and the PS1 public science archive have been made possible through contributions by the Institute for Astronomy, the University of Hawaii, the Pan-STARRS Project Office, the Max-Planck Society and its participating institutes, the Max Planck Institute for Astronomy, Heidelberg and the Max Planck Institute for Extraterrestrial Physics, Garching, The Johns Hopkins University, Durham University, the University of Edinburgh, the Queen's University Belfast, the Harvard-Smithsonian Center for Astrophysics, the Las Cumbres Observatory Global Telescope Network Incorporated, the National Central University of Taiwan, the Space Telescope Science Institute, the National Aeronautics and Space Administration under Grant No. NNX08AR22G issued through the Planetary Science Division of the NASA Science Mission Directorate, the National Science Foundation Grant No. AST-1238877, the University of Maryland, Eotvos Lorand University (ELTE), the Los Alamos National Laboratory, and the Gordon and Betty Moore Foundation.

AllWISE makes use of data from WISE, which is a joint project of the University of California, Los Angeles, and the Jet Propulsion Laboratory/California Institute of Technology, and NEOWISE, which is a project of the Jet Propulsion Laboratory/California Institute of Technology. WISE and NEOWISE are funded by the National Aeronautics and Space Administration.

The research leading to these results has received funding from the European Union Seventh Framework Programme FP7/2007-2013/ under grant agreement number 607254. This publication reflects only the author's view and the European Union is not responsible for any use that may be made of the information contained therein. KJD, HR and RJvW acknowledge support from the ERC Advanced Investigator programme NewClusters 321271. RJvW  acknowledges support from the VIDI research programme with project number 639.042.729, which is financed by the Netherlands Organisation for Scientific Research (NWO).

JS and PNB are grateful for support from the UK Science and Technology Facilities Council (STFC) via grant ST/M001229/1. WLW and MJH acknowledge support from the UK STFC [ST/M001008/1]. JHC and BM acknowledge support from the STFC under grants ST/M001326/1 and ST/R00109X/1. 
RKC and RK acknowledges support from the STFC through STFC studentships. 
LA acknowledges support from the STFC through a ScotDIST Intensive Data Science Scholarship.
LKM acknowledges the support of the Oxford Hintze Centre for Astrophysical Surveys which is funded through generous support from the Hintze Family Charitable Foundation. This publication arises from research partly funded by the John Fell Oxford University Press (OUP) Research Fund.

IP acknowledges support from INAF under PRIN SKA/CTA 'FORECaST'.
GJW gratefully acknowledges support from the Leverhulme Trust.
GGU acknowledges OCE Postocdoral fellowship from CSIRO.



\bibliographystyle{aa}
\bibliography{library}


\label{lastpage}
\end{document}